\documentclass[12pt,preprint]{aastex}
\usepackage{epsf}

\slugcomment{Submitted to the Astrophysical Journal - revised version}

\begin{document}

\title{THE $\rm \bf ^{12}C(\alpha,\gamma)^{16}O$ REACTION RATE 
AND THE EVOLUTION OF STARS IN THE MASS RANGE $\rm \bf 0.8 \leq M/M_{\odot} \leq 25$}

\author{Gianluca Imbriani\altaffilmark{1,6}, Marco Limongi\altaffilmark{2},
Lucio Gialanella\altaffilmark{1}, Filippo Terrasi\altaffilmark{3},
Oscar Straniero\altaffilmark{4} and Alessandro Chieffi\altaffilmark{5,2}}

\affil{1. Dipartimento di Scienze Fisiche, Universit\'a Federico II di Napoli, and INFN 
Napoli, Italy; imbriani@na.infn.it; lgialanella@na.infn.it} 

\affil{2. Osservatorio Astronomico di Roma, Via Frascati 33, I-00040, Monteporzio Catone, Italy; 
marco@nemo.mporzio.astro.it} 

\affil{3. Dipartimento di Scienze Ambientali, Seconda Universit\'a di Napoli, Caserta and INFN
Napoli, Italy; terrasi@na.infn.it} 

\affil{4. Osservatorio Astronomico di Collurania, I-64100, Teramo, Italy; straniero@astrte.te.astro.it}                   

\affil{5. Istituto di Astrofisica Spaziale (CNR), Via Fosso del Cavaliere, I-00133, Roma, Italy; 
achieffi@ias.rm.cnr.it} 

\affil{6. Osservatorio Astronomico di Napoli, via Moiariello 16, I-80131, Napoli, Italy}                   

\begin{abstract}
\scriptsize
We discuss the influence of the $\rm ^{12}C(\alpha,\gamma)^{16}O$ reaction rate on the central He 
burning of stars in the mass range $\rm 0.8-25~M_\odot$, as well as its effects on the
explosive yields of a $\rm 25~M_\odot$ star of solar chemical composition.
We find that the central He burning is only marginally affected by a change in this cross section within the currently accepted
uncertainty range. The only (important) quantity which varies significantly is the amount of C left by the He burning. 
Since the $\rm ^{12}C(\alpha,\gamma)^{16}O$ is efficient in a convective core,
we have also analyzed the influence of the convective mixing in determining the final C abundance left by the central He burning.
Our main finding is that the adopted mixing scheme does not influence the final C abundance provided the outer border
of the convective core remains essentially fixed (in mass) when the central He abundance drops below $\simeq 0.1$ dex by mass fraction;
vice versa, even a slight shift (in mass) of the border of the convective core during the last part of the central He burning
could appreciably alter the final C abundance.
Hence, we stress that it is wiser to discuss the advanced evolutionary phases as a function of the C abundance left by the He burning
rather than as a function of the efficiency of the $\rm ^{12}C(\alpha,\gamma)^{16}O$ reaction rate.
Only a better knowledge of this cross section and/or the physics of the convective motions could help in removing the degeneracy between
these two components.
We also prolonged the evolution of the two $\rm 25~M_\odot$ stellar models up to the core collapse and computed the final explosive yields. 
Our main results are that the intermediate-light elements, Ne, Na, Mg and Al (which are produced in the C 
convective shell) scale directly with the C abundance left by the He burning because they depend directly 
on the amount of available fuel (i.e C and or Ne). All the elements whose final yields are produced by any 
of the four explosive burnings (complete explosive Si burning, incomplete explosive Si burning, explosive 
O burning and explosive Ne burning) scale inversely with the C abundance left by the He burning because 
the mass-radius relation in the deep interior of a star steepens as the C abundance reduces.
We confirm previous findings according to which a low C abundance ($\simeq~0.2$ dex by mass fraction) 
is required to obtain yields with a scaled solar distribution.

\end{abstract}



\section{Introduction}\label{intro}

"The rate of the $\rm ^{12}C(\alpha,\gamma)^{16}O$ during hydrostatic helium burning 
is of vital interest for explosive nucleosynthesis. It is
this process that determines the abundances of $\rm ^{12}C$ and $\rm ^{16}O$ in the star, and thereby sets 
the stage for explosive burning...The rate is determined by the $\rm 7.115~ MeV$ level in the $\rm ^{16}O$ 
compound nucleus. At present the reduced width $\rm \theta_{\alpha}^2$ of this resonance for $\rm \alpha$ 
captures is {\it not known}." These sentences are taken from the 1973 issue of the
Ann.Rev.Astron.Astrophys. and were written by Arnett to emphasize both the
importance of this reaction in determining the
final yields produced by the explosion of a supernova event and the fact that
this rate was very uncertain. The experimental and theoretical efforts in the following 30 years led to
constrain the reduced width $\rm \theta_{\alpha}^2$ of the $\rm 7.115~MeV$ level in $\rm ^{16}O$ and therefore the
E1 component of the $\rm ^{12}C(\alpha,\gamma)^{16}O$ cross section. These studies, however, also
pointed out how other components
(not equally well constrained) contribute to the total cross section, so that the present determination
of the stellar rate of the $\rm ^{12}C(\alpha,\gamma)^{16}O$ is still affected by a large error.

From an experimental point of view, in
spite of the enormous efforts devoted
to the measurement of this cross section, the corresponding rate at
astrophysical energies is still far from being well established.
The cross section around the Gamow peak is dominated by ground state transitions through four different processes:
the two E1 amplitudes due to the low-energy tail of the $\rm 1^-$
resonance at $\rm E_{cm}=2.42~ MeV$ and to the subthreshold resonance at
$\rm -45~ keV$, the E2 amplitude due to the $\rm 2^+$ subthreshold
resonance at $\rm -245~ keV$ and the direct capture to the  $\rm ^{16}O$
ground state (plus the relevant interference terms).
Besides ground state transitions, also cascades, mainly through the E2 direct capture to the
$\rm 6.05~ MeV$ $0^+$ and $\rm 6.92~ MeV$ $2^+$ states, have to be considered. Although they are
believed to give a minor contribution (about 10 \%) to the total cross section, no experimental 
data are available for such transitions.
In the past twenty-five years many experiments have been
set-up, most of them based on the detection of $\rm \gamma$-rays
produced by $\rm \alpha$ captures in direct or inverse kinematics 
\citep{dy74, re85, re87, kr88, ou92, ou96, Ro99, gio}. 
All these measures extend to a minimum energy of about $\rm 1~ MeV$ and show systematic differences;
below this energy, the extremely
small value of the cross section ($\rm <10~pb$ ) hampers the direct
detection of $\rm \gamma$-rays and extrapolation procedures have to be
used in order to extract the astrophysical S-factor at the
relevant energies ($\rm E_0 = 300~ keV$ for $\rm T_9=0.18$). 
Such an extrapolation, which is based on the fitting of
differential cross sections in the investigated region, requires also the inclusion
of the phase correlation between the two incoming partial waves which contribute to
the two multipoles.

Additional information is provided by the elastic scattering data \citep{pl87} and by the $\rm \beta$-delayed 
$\rm \alpha$-decay of $^{16}N$ \citep{bu93,az94}.
Also the decay
to the first excited state has to be included together to a possible
non radiative E0 ground state transition. As
far as a consistent description of the E1 interference terms is concerned, it
should be noted that the evaluation of the contribution of higher energy $\rm 1^-$ levels
requires data to be taken at energies well above this
resonance, where the competition with the background arising from
the $\rm ^{13}C(\alpha,n)$ reaction (or other neutron-producing reactions
in inverse kinematics studies) makes cross section measurements
very difficult.

The above arguments make the extrapolated values of S(300) very
uncertain. A global analysis \citep{bu96} of all the
available data (surface fit) including the $\rm \gamma$ decay which follows an $\rm \alpha$ capture
from $\rm ^{12}C $, elastic scattering of $\rm \alpha$ particles from  $\rm ^{12}C $
and the $\rm \alpha$ emission which follows a $\rm \beta ^-$ decay of  $\rm ^{16}N$
\citep{bu93,az94} yielded a
wide range of results (from $\rm 62~ keV~ b$ to $\rm 270~ keV~ b$ ) for the
extrapolated S-factor. The minimum and maximum values which bracket such a spread correspond to reaction rates
(for $\rm T_9=0.18$) of $\rm 0.5\cdot 10^{-15}$ and
$\rm 2.2\cdot 10^{-15}~ cm^3/(mol\cdot s)$, which can be compared to the
data reported in the compilations of Caughlan and Fowler, 1988, ($\rm N_A\sigma v=0.8\cdot
10^{-15}~ cm^3/(mol\cdot s)$), hereinafter CF88, and Caughlan {\em et al.}, 1985, ($\rm N_A\sigma v=1.9\cdot
10^{-15}~ cm^3/(mol\cdot s)$), hereinafter CF85, which are generally used in stellar
evolution calculations. Finally, a recent compilation \citep{Na}
yields $\rm 0.9\cdot 10^{-15}~ cm^3/(mol\cdot s)$ and
$\rm 2.1\cdot 10^{-15}~ cm^3/(mol\cdot s)$ as lower and upper
bounds for this reaction rate, and $\rm 1.5\cdot 10^{-15}~ cm^3/(mol\cdot s)$ as the recommended value.

On the theoretical side, Arnett (1971) was the first to point out that the observed solar abundances of C and O could be
used to limit the rate of the $\rm ^{12}C(\alpha,\gamma)^{16}O$.
On the same line, Weaver and Woosley (1993) also tried to fix this rate by requiring the final explosive yields
to have a scaled solar relative distribution.

In addition to these efforts made to constrain this rate on the basis of the yields produced,
the direct influence of this process on the central He burning phase itself was also tested:
in particular, Iben (1968, 1972) and Brunish \& Becker (1990), by analyzing the behavior of a set of  
intermediate mass stellar models, found out that a change in the $\rm ^{12}C(\alpha,\gamma)^{16}O$ reaction rate
led to a change in the properties of the stars in the blue loop phase and hence, in turn, that it could modify
the mass range capable of entering the Cepheids instability strip. Contrarily to these 
results, Umeda et {\em al} 1999, Zoccali et {\em al} (2000) and Bono et {\em al} (2000)
found that a change in the $\rm ^{12}C(\alpha,\gamma)^{16}O$ rate does not modify
the path of a star in the HR diagram. 

For sake of completeness let us remind that also the properties of the
cooling sequences of the White Dwarfs have been studied as a function of the relative abundances of C and O in the He exhausted
core. We refer the reader to the papers
by, e.g., Segretain et {\em al} (1994), Salaris et {\em al} (1997), Brocato et {\em al} (1999) and Chabrier et {\em al} (2000) 
for an overview of the main findings in this field.

Though the partial effects of a change in this cross section on the evolution of a star 
have been addressed in several papers over the years (as we have already pointed out),
a comprehensive and homogeneous analysis of its effects over an extended mass interval is still missing.
Moreover, we believe that the interplay between the convection
and the $\rm ^{12}C(\alpha,\gamma)^{16}O$ in determining the 
chemical composition left by the He burning needs a deeper analysis.
In this paper we will analyze the dependence of the central He burning phase on the $\rm ^{12}C(\alpha,\gamma)^{16}O$ 
reaction rate over an extended mass interval, together with its interplay with the convective  mixing. 
We will also discuss the dependence of the final explosive yields produced by a $\rm 25~M_\odot$
on the C abundance left by the He burning.

The paper is organized as follows: in the next section we briefly remind the main properties of the
evolutionary code (FRANEC) adopted to perform all the computations. Section three is devoted
to the discussion of the central He burnings of all our models while the advanced
evolution of the $\rm 25~M_\odot$ stellar models are addressed in section four. The final explosive yields are presented in section five.
A final discussion and conclusion will follow.

\section{The evolutionary Code}

All the evolutionary tracks have been computed with the latest release (4.8) of the FRANEC (Frascati
RAphson Newton Evolutionary Code) whose earliest and latest version have been presented by Chieffi and
Straniero (1989) and Chieffi, Limongi and Straniero (1998). All the latest available input physics have
been adopted as discussed in Straniero, Chieffi and Limongi (1997). No mass loss has been taken into account.
The network adopted in the present set of models includes 19 isotopes for the evolution of the low
and intermediate mass stars and 179 isotopes for the evolutions of the $\rm 25~M_{\odot}$ stars.
Since we will discuss the effects of the overshooting and semiconvection on the stellar
models, and since for historical reasons these words have been used to mean
very different phenomena in stars of different mass, we briefly remind what they
refer to in the various mass ranges. 

\subsection{Overshooting and semiconvection in low mass stars}

During the central He burning, He is converted in C first and in O later. The increase of the C and
O abundances in the convective core raises the opacity so that a jump in the radiative gradient forms
at the border of the convective core. This is a condition of unstable equilibrium in the sense that
the possible mixing (by whichever phenomenon) of the radiative layers just outside the border of the convective core
would switch them from a stable to an unstable condition.
The reason is that the C brought in the radiative layer raises the radiative gradient (through the
opacity) so that it becomes intrinsically convective. This phenomenon,
usually called {\rm "induced" overshooting}, does not contain any 
free parameter which may be adjusted by hand since the process of "growth" of
the convective core is fully controlled by the requirement that the positive difference between the radiative and adiabatic
gradient cancels out. The word "induced" refers to the fact that this phenomenon is induced
by the conversion of He in C and O. When the central He abundance drops below $\simeq 0.6$ dex by mass fraction, the
radiative gradient does not decrease any more monotonically moving outward but it forms a minimum well inside the formal border
of the convective core. This occurrence triggers
the formation of a region (outside the mass location corresponding to this minimum) in which the matter is only partially mixed: 
the condition which controls the degree of mixing occurring in this region
is that the radiative gradient equals the adiabatic one (this equality is controlled, once again, by the opacity which,
in turn, depends on the local abundances of C and O in these layers). This is the so called "semiconvective" region
which forms in low mass stars.
For a much more detailed discussion of these phenomena we refer the reader to, e.g., 
Castellani et {\em al} (1985).
Since the "induced" overshooting and semiconvection completely depend
on the fact that the opacity is strongly dependent on the chemical composition, it is clear that they become progressively
less important, and eventually disappear, as the initial mass of a star increases because the electron scattering
(which does not depend on the chemical composition in an environment deprived of H) becomes the main source of the opacity.
In practice the semiconvective layer disappears for masses above $\rm \sim~5~M_{\odot}$ while the "induced" overshooting remains
at least partially efficient up to $\rm \sim ~ 20~M_{\odot}$.

The "real" existence of these phenomena in low mass stars is mainly supported 
by the star countings in the galactic globular clusters: in particular the ratio between the He burning stars (HB-stars) and
those ascending the Giant Branch (the first and/or the second time) can be explained only if the central He burning
timescale is the one obtained by including these two phenomena. Also in this case we refer the reader to
Castellani et {\em al} (1985) for a careful discussion of these problems.

During the latest part of the central He burning (i.e. when the He drops below 0.1 dex by mass fraction),
it has been recognized that a runaway of the outer border of the convective core occurs (usually called Breathing Pulse -hereinafter BP- , see
Castellani et {\em al}, 1985 and Caputo et {\em al}, 1989): its main effect is that of engulfing fresh He towards the center
and hence prolonging the central He burning lifetime. A discussion on the real existence of these instabilities is far beyond the purposes
of this paper (but see Castellani at al. 1985);
we simply want to stress the fact that their inclusion or suppression significantly alters also the
abundances of C and O at the end of the He burning.

\subsection{Overshooting and semiconvection in massive stars}
The word overshooting is used, in this case, to mean the phenomenon which would allow the convective bubbles
to penetrate the radiative layers surrounding a convective zone and hence to induce the mixing of a region 
larger (in mass) than
classically allowed by the strict adoption of the Schwarzschild criterion. This is a mechanical phenomenon which is
not confined to a specific evolutionary phase but which may be present at the border of any convective region. The
extension of this overshooted region is, in principle, totally arbitrary and usually
parameterized by imposing that the convective bubbles may reach a maximum extension over the formal convective
border that is proportional to the pressure scale
height ($\rm H_{p}$). The existence of a convective core larger than permitted by the Schwarzschild criterion
was invoked in the past 
in order to explain some observational data (see, e.g. Langer and Maeder, 1995,
but see also Testa et {\em al}, 1999). Though we do not intend to discuss here the possible existence or not
of a mechanical overshooting, it must be said that during the years the accepted size of this
phenomenon in the central H burning phase progressively reduced from $\rm \simeq1~H_p$ down to less than $\rm 0.2~H_p$.

The word semiconvection is used, in this framework, to mean the partial mixing which (would) occur at the end
of central H burning in the region of variable chemical composition left by the receding H-convective core
in stars more massive than $\rm \simeq 15 M_{\odot}$.
When the star exhausts the H in the center and readjusts on a structure supported by an H burning shell,
the radiative gradient overcomes the adiabatic one within these layers showing
a gradient of chemical composition. While these layers would be definitely convective
unstable if the Schwarzschild criterion were adopted to asses their stability, the adoption of the Ledoux criterion
would maintain these layers stable. Observational
constraints (see e.g. Langer and Maeder, 1995), mainly related to the observed number ratio between red and blue supergiants,
seem to favor the Ledoux criterion, i.e. a partial or even negligible amount of mixing.
 
Before closing this section let us clearly state that our {\rm standard} computations are obtained by adopting always
the Schwarzschild criterion but in the central He burning phase where both the "induced" overshooting and semiconvection
are properly taken into account while the BPs are quenched by forcing the central He abundance to be a monotonic not 
increasing function of time. No mechanical overshooting has been included. Moreover, no mixing
is allowed in the semiconvective H-rich layers (which corresponds to a strict application of the Ledoux criterium). 

\section{The central Helium Burning phase}
We followed the evolution of stellar model having $\rm 2.5\leq M/M_{\odot}\leq 25$, Y=0.285 and Z=0.02
from the Main Sequence up to the central He exhaustion.
We also followed the central He burning phase of a typical Globular Cluster Horizontal Branch (HB) star, i.e. a star with
an He core mass of $\rm 0.485~M_{\odot}$, a total mass of $\rm 0.6~M_{\odot}$, an initial He abundance Y=0.23 and a metallicity
Z=0.001. All these evolutions have been computed twice: firstly by adopting the
$\rm ^{12}C(\alpha,\gamma)^{16}O$ reaction rate provided by CF88 and secondly by adopting the one provided by CF85. 

\begin{deluxetable}{cccccccccccccccc}
\label{tab_He}\tabletypesize{\scriptsize} 
\tablewidth{0pt}
\tablecaption{Main properties of the central Helium burning phase.}
\tablehead{
\colhead{$\bf M$}              &    
\multicolumn{2}{c}{$\bf t_{He}$}         &
\multicolumn{2}{c}{$\bf X_{^{12}C}$}     &
\multicolumn{2}{c}{$\bf X_{^{16}O}$}     &
\multicolumn{2}{c}{$\bf t_B$ }           &
\multicolumn{2}{c}{$\bf t_B/t_{He}$}     &
\colhead{$\bf M^{1}_{CHe}$}    &
\multicolumn{2}{c}{$\bf M_{CC}$}         &
\multicolumn{2}{c}{$\bf M^2_{CHe}$ }   \\
\colhead{$\bf (M_\odot)$} &    
\multicolumn{2}{c}{$\bf (Myr) $ }   &
\multicolumn{2}{c}{} &
\multicolumn{2}{c}{} &
\multicolumn{2}{c}{$\bf (Myr)$ }    &
\multicolumn{2}{c}{$\bf (\%)$  }    &
\colhead{$\bf (M_\odot)$} &
\multicolumn{2}{c}{$\bf (M_\odot)$} &
\multicolumn{2}{c}{$\bf (M_\odot)$} \\
\colhead{}               &
\colhead{88}           &
\colhead{85}           &
\colhead{88}           &
\colhead{85}           &
\colhead{88}           &
\colhead{85}           &
\colhead{88}           &
\colhead{85}           &
\colhead{88}           &
\colhead{85}           &
\colhead{}            &
\colhead{88}           &
\colhead{85}           &
\colhead{88}           &
\colhead{85}
                  }
\startdata
$ 0.8$      &    $100$  & $110$    &      $0.495$  &  $0.294$    &     $0.505$  &    $0.786$     &     $0.00  $  &   $0.00$    &  $0   $   &   $0 $  &   $0.48$   & $0.200$   & $0.220$    &    $0.50$   & $0.50$  \\ 
$ 2.5$      &    $219$  & $231$    &      $0.451$  &  $0.195$    &     $0.530$  &    $0.761$     &     $0.00  $  &   $0.00$    &  $0   $   &   $0$   &   $0.33$   & $0.200$   & $0.210$    &    $0.50$   & $0.51$  \\ 
$ 3$        &    $124$  & $134$    &      $0.493$  &  $0.221$    &     $0.489$  &    $0.691$     &     $0.00  $  &   $0.00$    &  $0   $   &   $0$   &   $0.38$   & $0.223$   & $0.224$    &    $0.55$   & $0.56$  \\ 
$ 5$        &    $18.9$ & $20.7$   &      $0.556$  &  $0.290$    &     $0.425$  &    $0.688$     &     $0.00  $  &   $0.00$    &  $0   $   &   $0$   &   $0.64$   & $0.433$   & $0.452$    &    $1.03$   & $1.05$  \\ 
$ 6$        &    $10.9$ & $11.7$   &      $0.541$  &  $0.294$    &     $0.440$  &    $0.711$     &     $4.88  $  &   $5.76$    &  $45  $   &   $49$  &   $0.80$   & $0.552$   & $0.571$    &    $1.31$   & $1.34$  \\ 
$ 8$        &    $4.98$ & $5.45$   &      $0.524$  &  $0.270$    &     $0.457$  &    $0.736$     &     $2.50  $  &   $2.99$    &  $50  $   &   $55$  &   $1.16$   & $0.818$   & $0.878$    &    $1.90$   & $1.94$  \\                                                                
$ 10$       &    $2.87$ & $3.15$   &      $0.501$  &  $0.245$    &     $0.480$  &    $0.744$     &     $1.37  $  &   $1.66$    &  $47  $   &   $53$  &   $1.58$   & $1.152$   & $1.231$    &    $2.52$   & $2.58$  \\                      
$ 12$       &    $1.88$ & $2.07$   &      $0.490$  &  $0.237$    &     $0.492$  &    $0.752$     &     $0.76  $  &   $0.97$    &  $40  $   &   $47$  &   $2.08$   & $1.555$   & $1.576$    &    $3.19$   & $3.25$  \\ 
$ 14$       &    $1.36$ & $1.50$   &      $0.482$  &  $0.229$    &     $0.499$  &    $0.751$     &     $0.33  $  &   $0.57$    &  $24  $   &   $38$  &   $2.65$   & $2.006$   & $2.126$    &    $3.89$   & $3.97$  \\ 
$ 15$       &    $1.20$ & $1.31$   &      $0.480$  &  $0.230$    &     $0.501$  &    $0.765$     &     $0.00  $  &   $0.37$    &  $0   $   &   $29$  &   $2.96$   & $2.271$   & $2.362$    &    $4.17$   & $4.33$  \\ 
$ 20$       &    $0.76$ & $0.83$   &      $0.453$  &  $0.216$    &     $0.527$  &    $0.794$     &     $0.00  $  &   $0.00$    &  $0   $   &   $0$   &   $4.66$   & $3.890$   & $4.041$    &    $6.33$   & $6.33$  \\ 
$ 25$       &    $0.58$ & $0.64$   &      $0.417$  &  $0.184$    &     $0.562$  &    $0.562$     &     $0.00  $  &   $0.00$    &  $0   $   &   $0$   &   $6.63$   & $5.862$   & $6.000$    &    $8.68$   & $8.68$
\enddata                                                                                                              
\end{deluxetable}                                                                               

Table \ref{tab_He} summarizes, for each mass, 
the main evolutionary properties in rows 1 to 12 (each couple of columns refers to the values
obtained with the CF88 and the CF85 rates, respectively). In order, left to right we report:
the central He burning lifetime, the C and O abundances left by the He burning,
the time spent by each model in the blue
loop (i.e. at $\rm Log(T_{eff}\geq3.80$)), the fraction of the He burning lifetime spent in the blue loop,
the He core mass at the He ignition, the maximum size of the convective core and the final
He core mass at the He exhaustion.

\begin{figure}

\vspace*{0cm}
\mbox{ \epsfxsize=\linewidth
       \epsffile{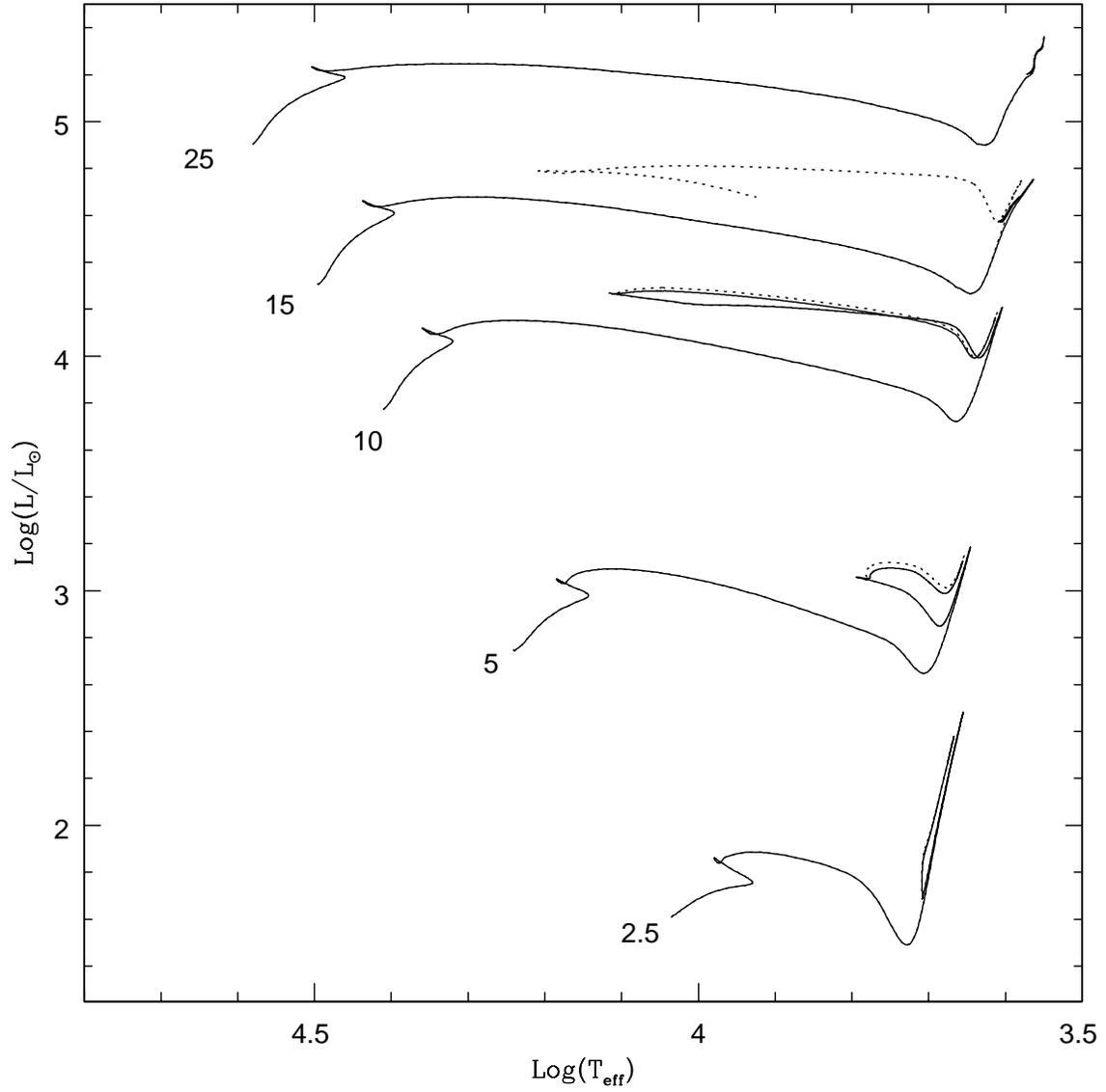}
     } 

\figcaption[f1.eps]{Path followed by selected models in the HR diagram; the dotted and solid lines refer, respectively, 
                       to the CF85 and CF88 cases.\label{HR}}

\end{figure}

\begin{figure}

\vspace*{0cm}
\mbox{ \epsfxsize=\linewidth
       \epsffile{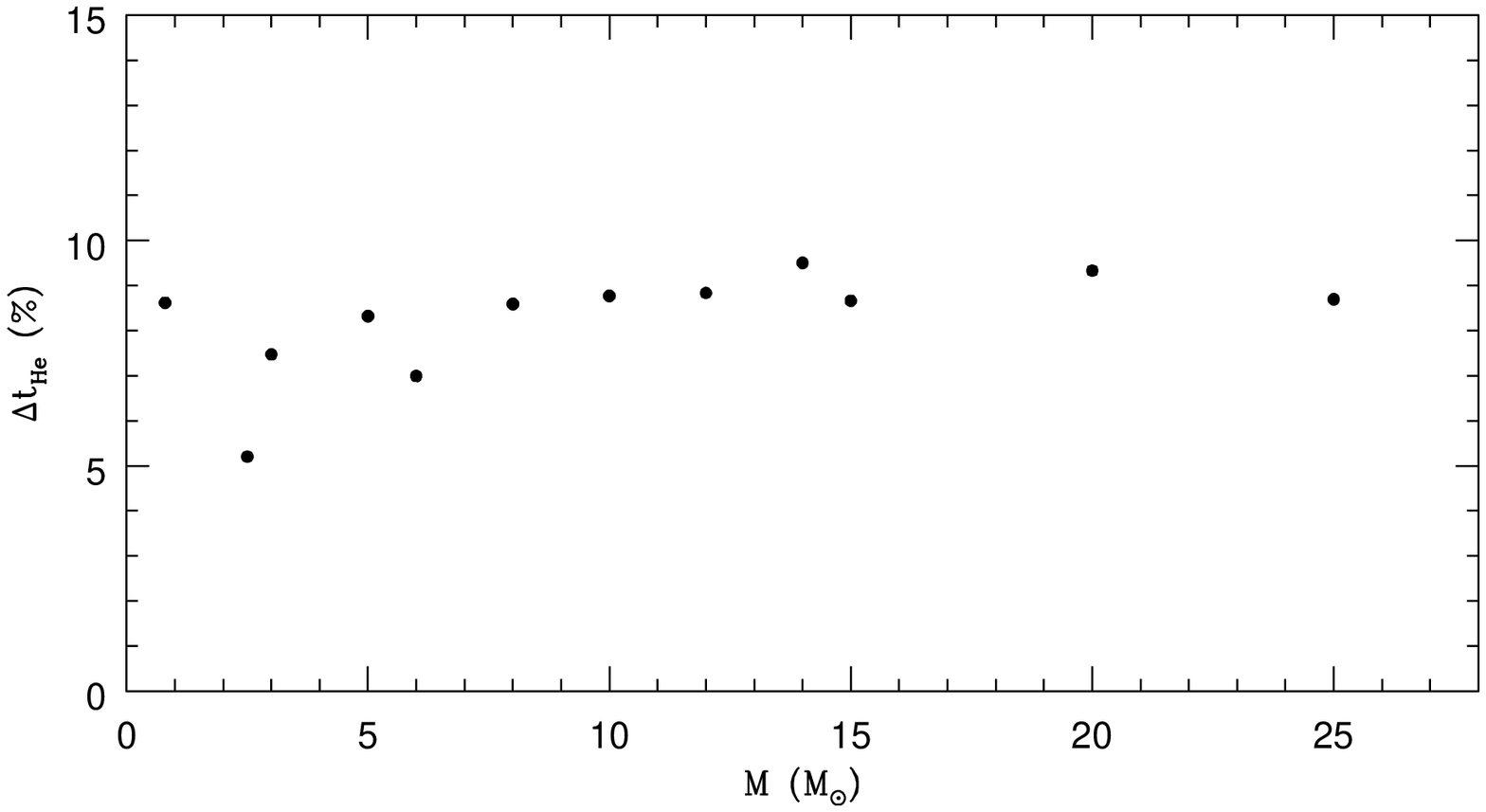}
     } 

\figcaption[f2.eps]{Dependence of the helium burning lifetime on the $\rm ^{12}C(\alpha,\gamma)^{16}O$ reaction rate 
                       as a function of the initial mass; 
                       each dot corresponds to $\rm (t_{He}^{CF85}-t_{He}^{CF88})/t_{He}^{CF85}$.\label{t_He}}

\end{figure}

\begin{figure}

\vspace*{0cm}
\mbox{ \epsfxsize=\linewidth
       \epsffile{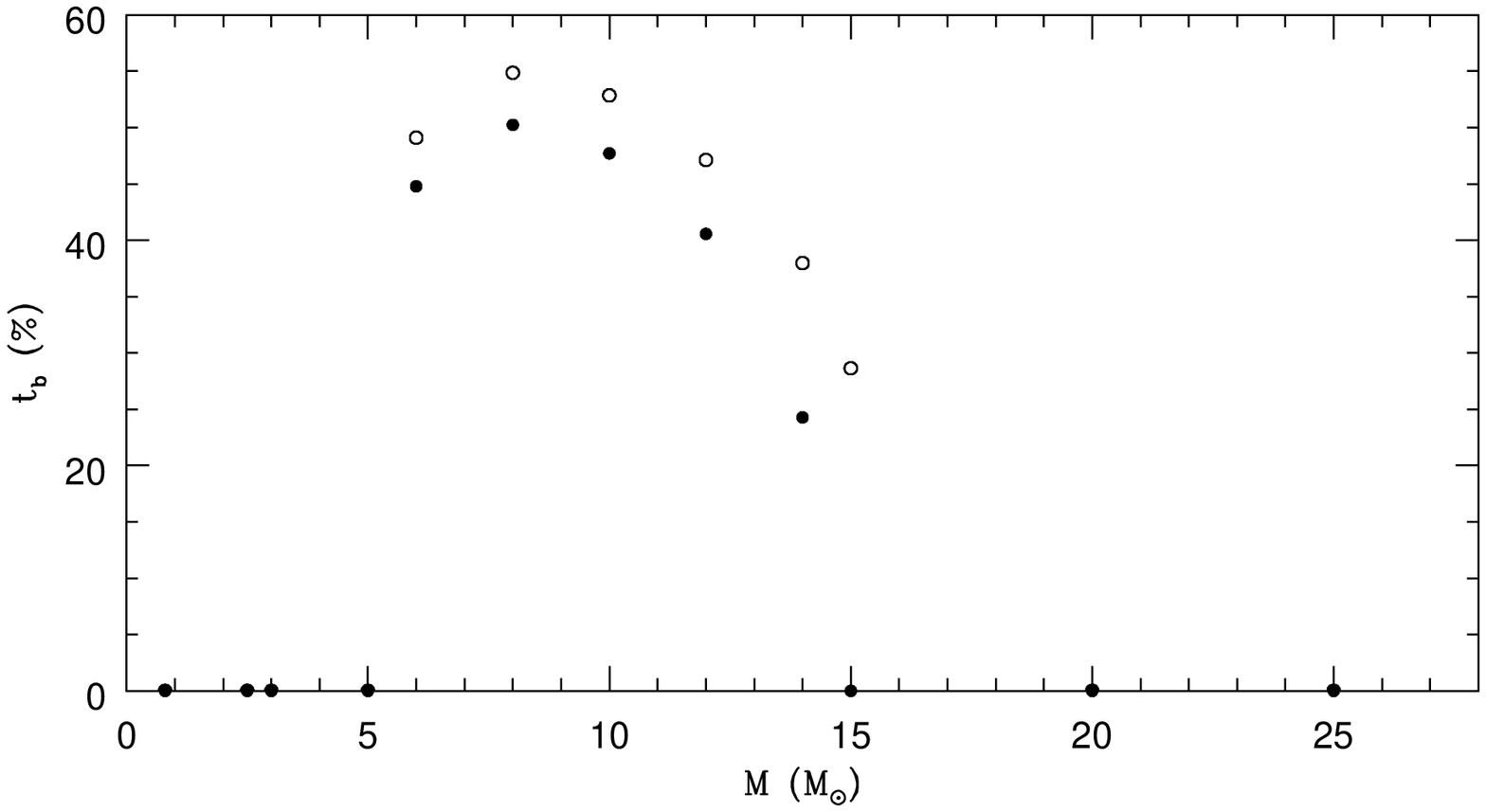}
     } 

\figcaption[f3.eps]{Dependence of the "blue loop lifetime" on the $\rm ^{12}C(\alpha,\gamma)^{16}O$ reaction rate 
                       as a function of the initial mass; each dot shows the percentage of the helium burning lifetime
                       spent at $\rm \log T_{eff}\ge 3.8$; 
                       the open and filled dots refer, respectively, to the CF85 and CF88
                       cases. \label{t_B}}

\end{figure}

Figures \ref{HR} to \ref{t_B} graphically show the effect of a change in
the $\rm ^{12}C(\alpha,\gamma)^{16}O$ reaction rate from the CF88 to the CF85 one on the central He burning phase.

In particular:
Figure \ref{HR} shows the path followed by a selected sample of stellar masses in the HR
diagram (the solid and dashed lines refer, respectively, to models computed with the CF88 and CF85 rate),
Figure \ref{t_He} shows, as filled dots, the difference in the He burning lifetimes (in percentage) obtained for
the two rates as a function of the initial mass; Figure \ref{t_B} shows, instead, the percentage of the He burning lifetime 
spent in the blue loop
(the filled and open dots refer, respectively, to models computed with the CF88 and CF85 rate).
All these three figures show that an uncertainty of the $\rm ^{12}C(\alpha,\gamma)^{16}O$ within the quoted range
does not dramatically alter the "observable" properties of a star in the central He burning phase.
In particular, the path followed by these stars in the HR diagram is practically unaffected by such a change
whereas both the total He burning lifetime and the time spent in the blue loop change by 10\% at most.
It is worth noting that the most massive star which experiences a blue loop in the central He burning phase 
changes from the 14 to 15 $\rm M_{\odot}$ as a consequence of the quoted change in  
the $\rm ^{12}C(\alpha,\gamma)^{16}O$ rate.

\begin{figure}

\vspace*{0cm}
\mbox{ \epsfxsize=\linewidth
       \epsffile{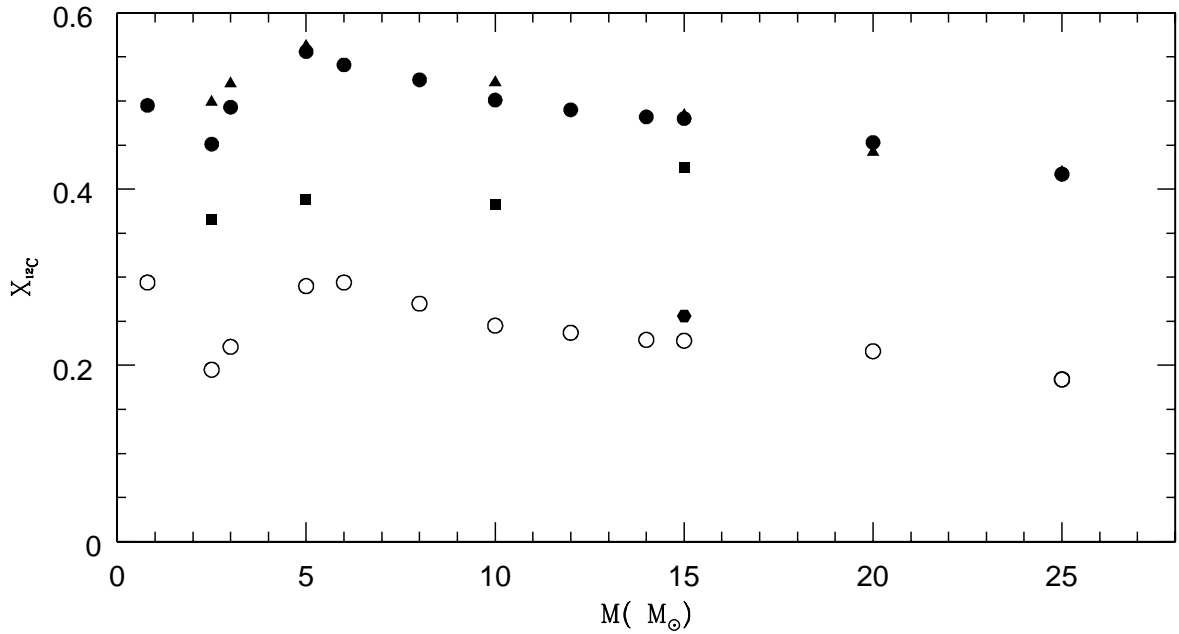}
     } 

\figcaption[f4.eps]{Carbon abundance left by the helium burning; the open and filled symbols refer, respectively, 
                       to the CF85 and CF88 cases. The dots represent our standard computations; the triangles represent
                       the runs obtained by including $\rm 1 ~H_p$ of mechanical overshooting; the squares refer to
                       the runs obtained by including the breathing pulses; the hexagon refers to the run in which 
                       $\rm 0.1~H_p$ of mechanical overshooting is imposed when the central helium abundance drops below
                       $\rm 0.075$ dex by mass fraction. \label{12C}}

\end{figure}

Let us turn now to the chemical composition left by the He burning; Figure \ref{12C} shows the amount of C left
by the He burning as a function of the initial mass.

The filled symbols always refer to computations performed by adopting the CF88 value while the open symbols 
always refer to models computed by adopting the CF85
value; the dots refer to our "standard" models. The first thing worth noting is that
the two sets of models show essentially the same dependence of the final C abundance on the
initial mass and hence they are more or less
systematically shifted one with respect to the other by 0.20:0.25 dex.
The general trend is that the  
C abundance left by the He burning increases as the initial mass reduces;
a maximum is then reached for a mass of the order of 5 $\rm M_{\odot}$ and then a drop
occurs for smaller values of the mass; the HB star 
behaves almost like the $\rm 2.5~M_{\odot}$.
It has to be noted that the maximum variation of the C abundance is of the order of 0.15 dex over
the full mass range under exam, and that this variation even reduces to 0.1 dex for the masses larger than
$\rm \simeq 8~M_{\odot}$. 
The existence of a smooth monotonic trend for masses larger than $\rm 5~M_\odot$ can be understood by reminding that 
a smaller mass favors the C production rather than its destruction because the
$\rm 3\alpha$ reaction rate scales with the square of the density while the $\rm ^{12}C(\alpha,\gamma)^{16}O$ scales
linearly with the density. The inversion of the trend for masses smaller than $\rm \simeq 5~M_\odot$
is probably due to the fact that stars with a very small He core mass spend enough time in the last part
of the central He burning so that the conversion of C in O is strongly favored.
The HB star behaves like the $\rm 2.5~M_\odot$ because they have a similar He core mass.

\begin{figure}

\vspace*{0cm}
\mbox{ \epsfxsize=\linewidth
       \epsffile{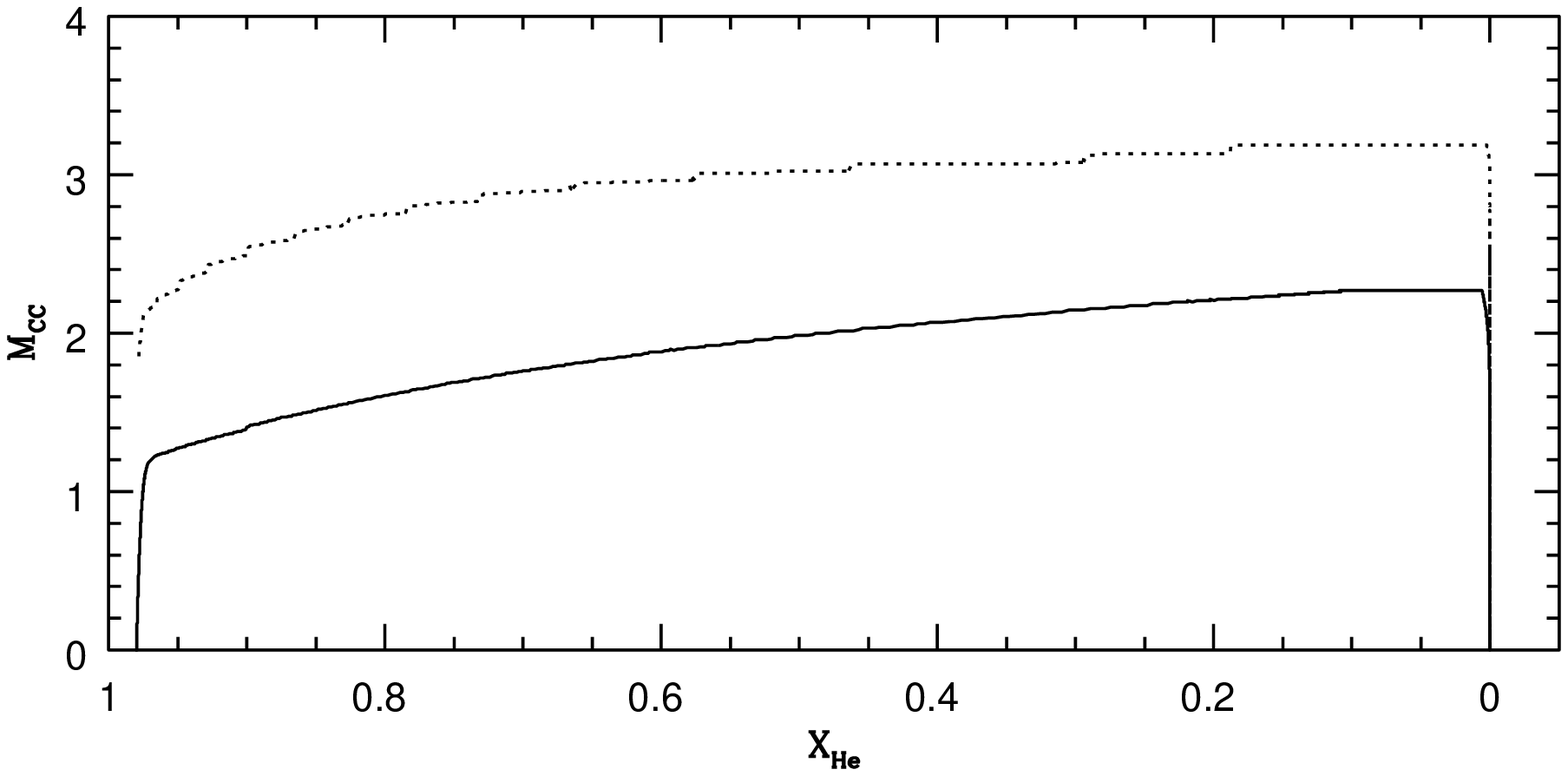}
     } 
\figcaption[f5.eps]{Template behavior of the border of the convective core versus the central He abundance 
                       for the present set of models. The solid line refers to a standard model while the dotted line refers to the
                       same model but computed by adding $\rm 1~H_p$ of overshooting.\label{CC}}

\end{figure}

Since the process we are dealing with occurs in a convective environment it is important to verify if, and
to what extent, the final C abundance depends on the adopted convective scenario.
Let us start with the standard one, i.e. the case in which the stability is controlled by the Schwarzschild criterion.
Figure \ref{CC} shows, as a solid line, the typical behavior of the convective core as a function of the central He abundance.
This figure shows that the convective core grows during the first part of the central He burning phase, reaches an asymptotic
value and then remains constant (in mass) until the possible occurrence of the BP (if ${\rm M\leq 15~M_\odot}$)
or till the end of the central He burning (masses above the ${\rm 15~M_\odot}$ never develop BPs). Since in our standard scenario the
BP are quenched out, Figure \ref{CC} represents the qualitative behavior of the convective core of all the stars in the mass interval
here studied. As we have already mentioned above, the final C abundance which is obtained by adopting these assumptions
is shown as filled dots in Figure \ref{12C}.

\begin{figure}

\vspace*{0cm}
\mbox{ \epsfxsize=\linewidth
       \epsffile{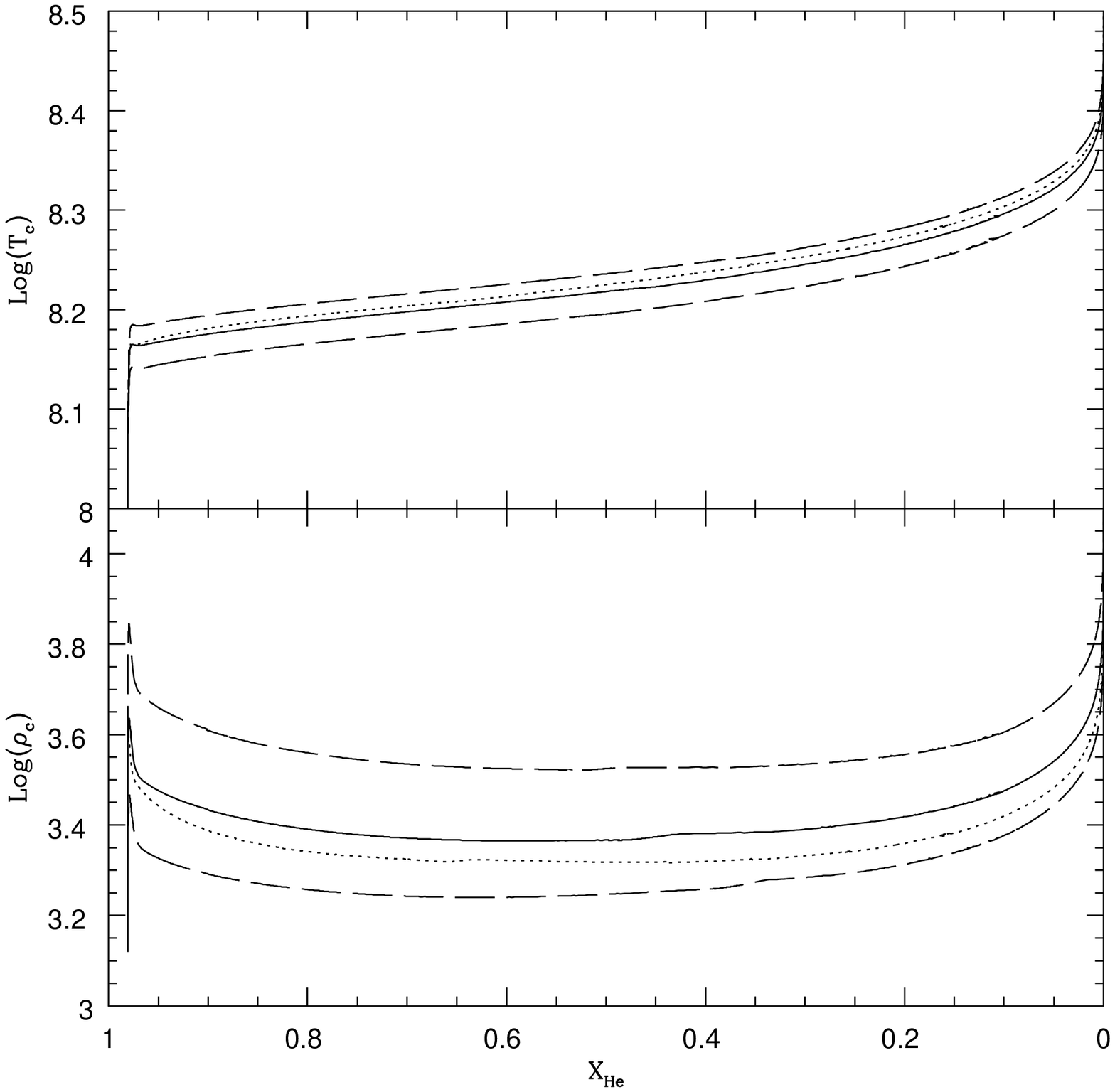}
     } 
\figcaption[f6.eps]{Run of the central temperature (upper panel) and density (lower panel) as a function of the central
                       He abundance. The solid line refers to the standard 10 $\rm M_{\odot}$, while the dotted one refers
                       to the 10 $\rm M_{\odot}$ computed with $\rm 1~H_p$ of mechanical overshooting. The two dashed lines show,
                       as a reference, the behavior of a 12 $\rm M_{\odot}$ and an 8 $\rm M_{\odot}$.
                       All these models were computed by adopting the $\rm ^{12}C(\alpha,\gamma)^{16}O$ reaction
                       rate as provided by CF88.\label{comp}}

\end{figure}

A second set of models spanning essentially the same mass interval has been recomputed by imposing a large overshooting 
($\rm 1~H_p$) during the central He burning phase. For sake of clarity let us remind the reader that the adoption of a large amount
of overshooting automatically cancels out the possible formation of a semiconvective region because it completely mixes the region 
where the partial mixing should occur.
In spite of a much larger (mass) size of the convective core (dotted line in Figure \ref{CC}), the C abundance left by the He burning
(shown as filled triangles in Figure \ref{12C}) closely resembles the one obtained in the standard case.
The reason is that the overshooting increases the size of the convective
core but it does not alter the behavior of the border of the convective core, which remains essentially constant in mass during the
latest part of the He burning; hence the run of both the central temperature and density as a function of the 
He abundance (see Figure \ref{comp}) do not change significantly, as well as
the rate at which He is converted in C, and C in O. The only effect of the overshooting is, in this respect, to increase
the He burning lifetime as a consequence of the increased amount of available fuel.

This picture changes drastically if one allows the border of the convective core to grow in mass when the central He drops below
$\rm \simeq 0.1$ dex by mass fraction. This possibility may "naturally" occur if, e.g., one did not artificially dump out the
occurrence of the BPs. The ingestion of fresh He in an environment very $\rm ^{12}C$ rich would favor, in this case, the 
$\rm ^{12}C(\alpha,\gamma)^{16}O$ rather than the $\rm 3\alpha$'s,
so that the final C abundance would be much lower than in the previous two scenarios. 
Moreover, since the number and the strength of the BPs scale inversely with the initial mass, it is clear that the lower
the mass the larger will be their influence on the evolution of the star.

The same effect which is obtained by means of the BPs may be obviously obtained in all cases in which even a small amount of
fresh He is allowed to enter the convective core towards the end of the central He burning.
In order to further stress how delicate the dependence of the final C abundance is on the behavior of the border of the convective
core in the latest phases of the He burning, we show
as a filled hexagon in Figure \ref{12C}, the
C abundance left by the He burning of a 15 $\rm M_{\odot}$ in which just 0.1 $\rm H_p$ of mechanical overshoot is
imposed when the central He burning drops below 0.075 dex by mass fraction: in this case the final C abundance even
resembles the value obtained by adopting the CF85 rate.

Before closing this section let us remark that, since massive stars do not have BPs
and since the size of the convective core does not alter the final C abundance at the end of the He burning, one could
be tempted to conclude that the C abundance left by the He burning depends only on the adopted value of the
$\rm ^{12}C(\alpha,\gamma)^{16}O$ rate; however, since we do not feel confident to state that current uncertainties 
in the treatment of the convective core of the massive stars
are merely confined to the size of the convective region itself, we prefer to conclude that in all the mass interval 
under exam the final C abundance left by the He burning depends on both the mixing scheme and
the adopted $\rm ^{12}C(\alpha,\gamma)^{16}O$ reaction rate.

\section{The advanced evolutionary phases of a $\rm 25~M_{\odot}$}

In the previous section we have shown the direct influence of the $\rm ^{12}C(\alpha,\gamma)^{16}O$ process on the central He burning of stars
in a wide mass interval together with its interplay with the treatment of the convective core.
The next logical step would be to follow the further evolution of all these stars in order to determine the final
impact of this process on stars of different masses. Such a big project goes beyond the purposes of the present paper:
in this section we will concentrate on the further evolution of the $\rm 25~M_\odot$ (taken as representative of the massive stars)
up to the final collapse and explosion. The C abundance left by the He burning is $\rm \simeq~0.4$ for the CF88 rate
and $\rm \simeq~0.2$ for the CF85 one. Since all the evolutionary 
properties discussed below depend directly on the C abundance left by the He burning but not (necessarily)
directly on the adopted value for the $\rm ^{12}C(\alpha,\gamma)^{16}O$ rate (see the previous section),
we think that the two runs obtained by adopting the CF88
and the CF85 rates must be discussed in terms of the C abundance left by the He burning
($\rm C_{\rm ini}$).
For this reason in this and the following sections we will change terminology: 
the run computed by adopting the CF88 rate will be
referred to as the $\rm C_{\rm 0.4}$ case, to underline that the results directly depend on a C abundance equal to 0.424.
Analogously, the run obtained by adopting the CF85 rate will be referred to as the ${\rm C_{\rm 0.2}}$ case.

\begin{deluxetable}{lclc}
\tabletypesize{\scriptsize} 
\label{tab_25}
\tablewidth{0pt}
\tablecaption{Main stages of the two $25M_\odot$ stars.}
\tablehead{
\colhead{} &
\colhead{$\rm CF85$} &
\colhead{} &\colhead{$\rm CF88$}
}
\startdata

&&\bf{H Burning}&  \\  \hline

$\tau_H (yr)$                     &   5.81(6) & &   5.81(6)  \\   
$M_{CC} (M_\odot)$                &   12.7     & &   12.7     \\ \hline
        
&&\bf{He Burning}&  \\  \hline        
        
$\Delta$t(H-exh.He-ign.)          &   2.70(4) & &   2.70(4) \\
$\tau_He (yr)$                    &   5.8(5)  & &   6.37(5) \\   
$M_{CC} (M_\odot)$                &   5.6     & &   5.8     \\ 
$\Delta t_{He ~conv~shell}$(yr)    &   1.6(4)  & &   1.5(4) \\
$\Delta M_{He ~conv~shell}(M\odot)$&  2.1      & &   2.2    \\
$^{12}C$                          &  0.424    & &   0.200  \\ 
$^{16}O$                          &  0.546    & &   0.769  \\ \hline
        
&&\bf{C Burning}&  \\  \hline        
        
$\Delta$t(He-exh.C-ign.)           &   1.17(4) & &   1.03(4) \\
$\tau_C (yr)$                      &   5.76(3) & &   4.56(3) \\   
$M_{CC} (M_\odot)$                 &    0.5    & &          \\
$\Delta t_{1 C ~conv~shell}$(yr)    &     91    & &    1     \\
$\Delta M_{1 C ~conv~shell}(M\odot)$&      1    & &    1.2   \\
$\Delta t_{2 C ~conv~shell}$(yr)    &     40    & &    0.2    \\
$\Delta M_{2 C ~conv~shell}(M\odot)$&      3    & &    2.4   \\
         
$^{16}O$                           &   0.378   & &   0.674   \\
$^{20}Ne$                          &   0.478   & &   0.260   \\ 
$^{24}Mg$                          &   0.014   & &   0.076   \\ \hline
        
&&\bf{Ne Burning}&  \\  \hline        
        
$\tau_Ne (yr)$                     &   37.9    & &   6.01    \\   
$M_{CC} (M_\odot)$                 &   0.56    & &   0.77    \\ 
$^{16}O$                           &   0.632   & &   0.810   \\ 
$^{24}Mg$                          &   0.139   & &   0.072   \\
$^{28}Si$                          &   0.143   & &   0.071  \\ \hline
        
&&\bf{O Burning}&  \\  \hline         
        
$\tau_O (yr)$                      &   1.62   & &   0.274     \\   
$M_{CC} (M_\odot)$                 &   1.26   & &   0.98    \\ 
$^{28}Si$                          &   0.561  & &   0.604  \\ 
$^{32}S$                           &   0.014  & &   0.008  \\ 
$^{34}S$                           &   0.336  & &   0.150  \\ \hline
        
&&\bf{Si Burning}&  \\  \hline         
        
$\tau_{Si} (yr)$                     &   0.21   & &   0.0167     \\   
$M_{CC} (M_\odot)$                 &   0.95   & &   1.28   \\ 
$^{56}Fe$                          &   0.507  & &   0.674  \\ 
$^{60}Ni$                          &   0.007  & &   0.023  
\enddata
\end{deluxetable}

\begin{figure}

\vspace*{0cm}
\mbox{ \epsfxsize=\linewidth
       \epsffile{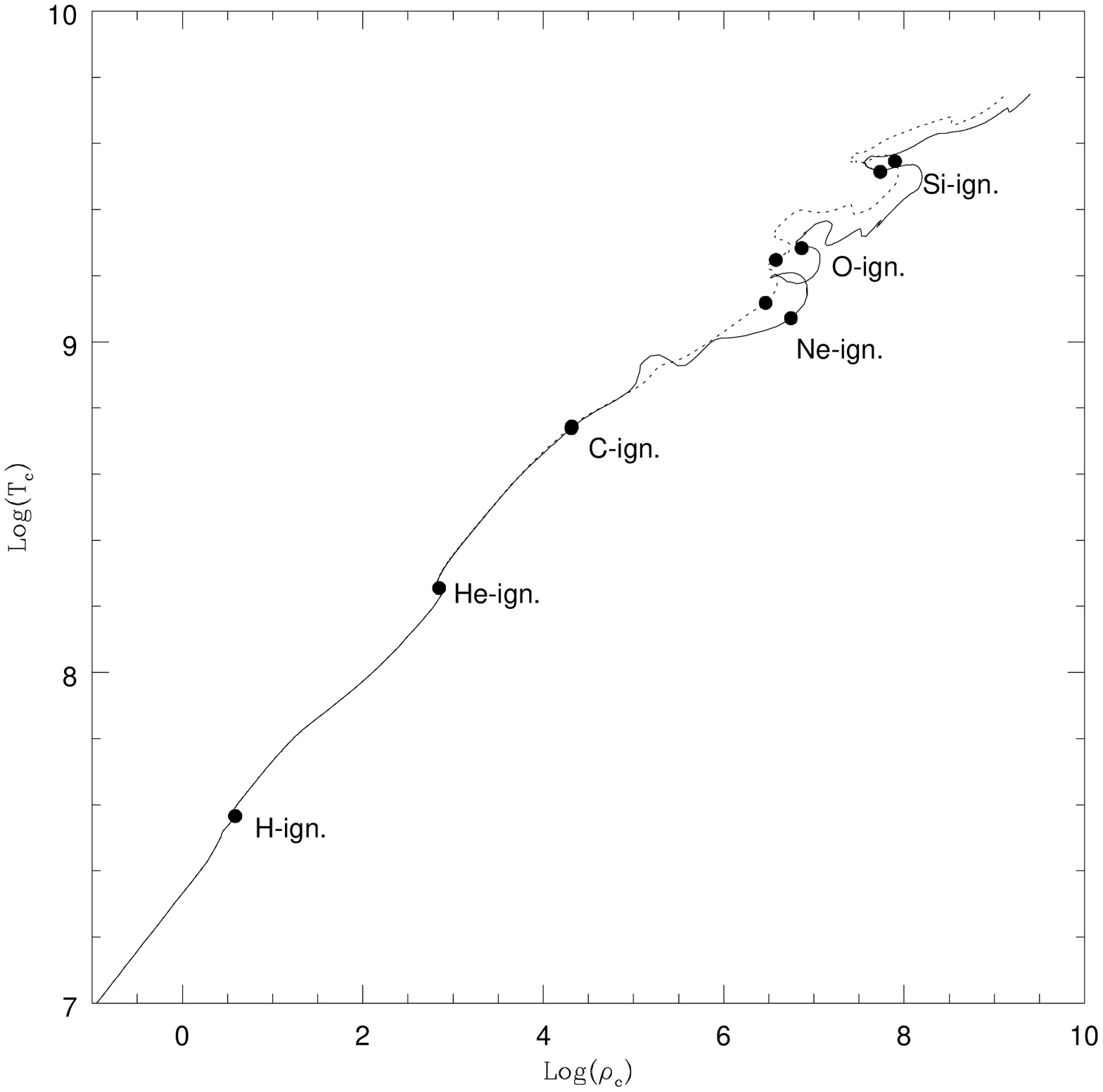}
     } 
\figcaption[f7.eps]{Run of the two $\rm 25~M_\odot$ in the $\rm Log(T_c) \div Log(\rho_c)$ plane. The solid and dotted lines
                       refer, respectively, to the $\rm C_{0.4}$  and $\rm C_{0.2}$ cases. See text.\label{tcro}}

\end{figure}

\begin{figure}

\vspace*{0cm}
\mbox{ \epsfxsize=\linewidth
       \epsffile{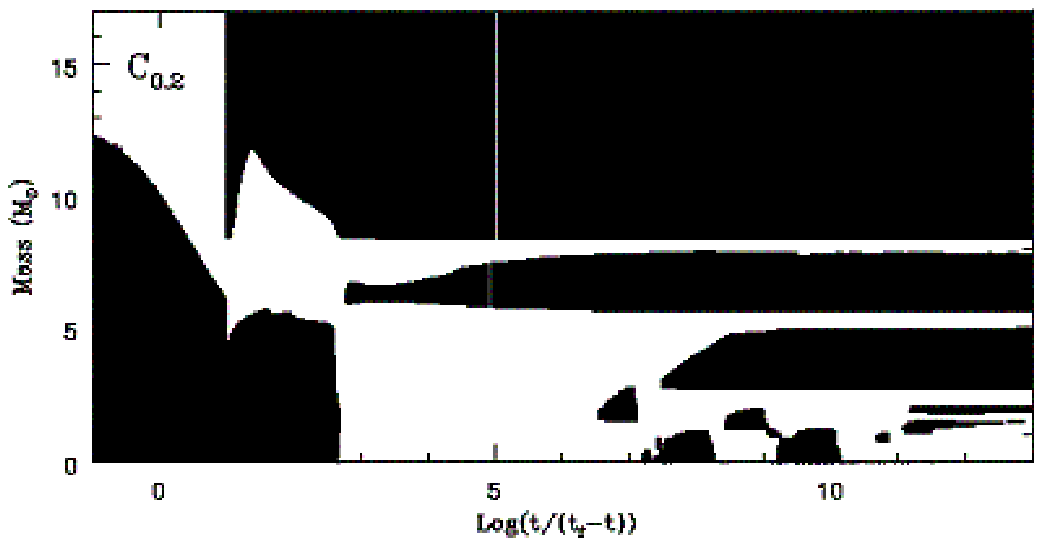}
     } 

\figcaption[f8.eps]{Temporal (properly adapted) behavior of the convective zones which form during the evolution of
                       the $\rm 25~M_\odot$ stellar model in the $\rm C_{0.2}$ case.\label{conv1}}

\end{figure}

\begin{figure}

\vspace*{0cm}
\mbox{ \epsfxsize=\linewidth
       \epsffile{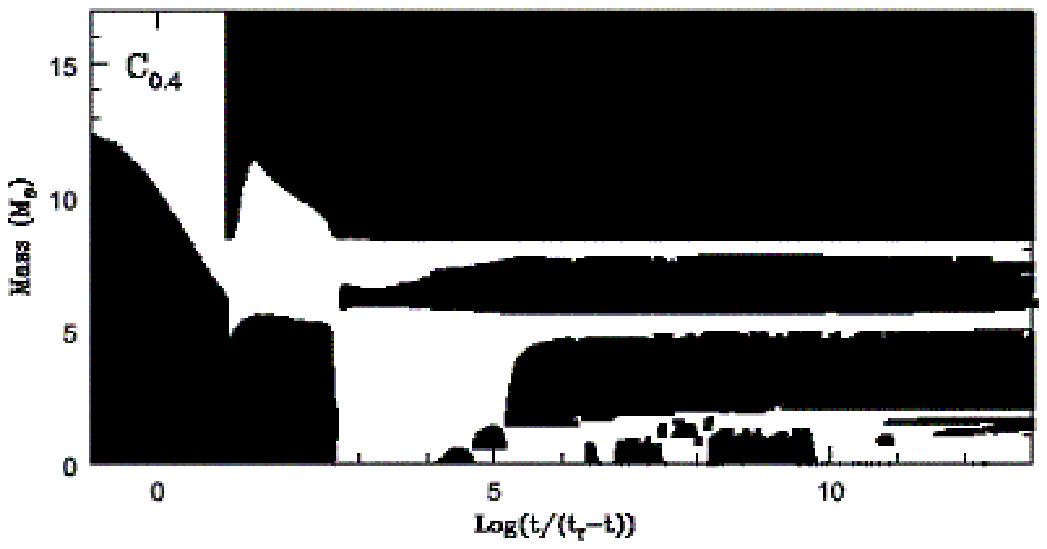}
     } 
\figcaption[f9.eps]{Temporal (properly adapted) behavior of the convective zones which form during the evolution of
                       the $\rm 25~M_\odot$ stellar model in the $\rm C_{0.4}$ case.\label{conv2}}

\end{figure}

The main evolutionary properties of these two evolutions are summarized in table \ref{tab_25} and 
in Figures \ref{tcro}, \ref{conv1} and \ref{conv2}.
Table \ref{tab_25} reports, for each central burning, its lifetime, the size of the convective core, the abundance of the most 
abundant elements produced in the burning as well as the data relative to the convective shell episodes, if present.

\begin{figure}

\vspace*{0cm}
\mbox{ \epsfxsize=\linewidth
       \epsffile{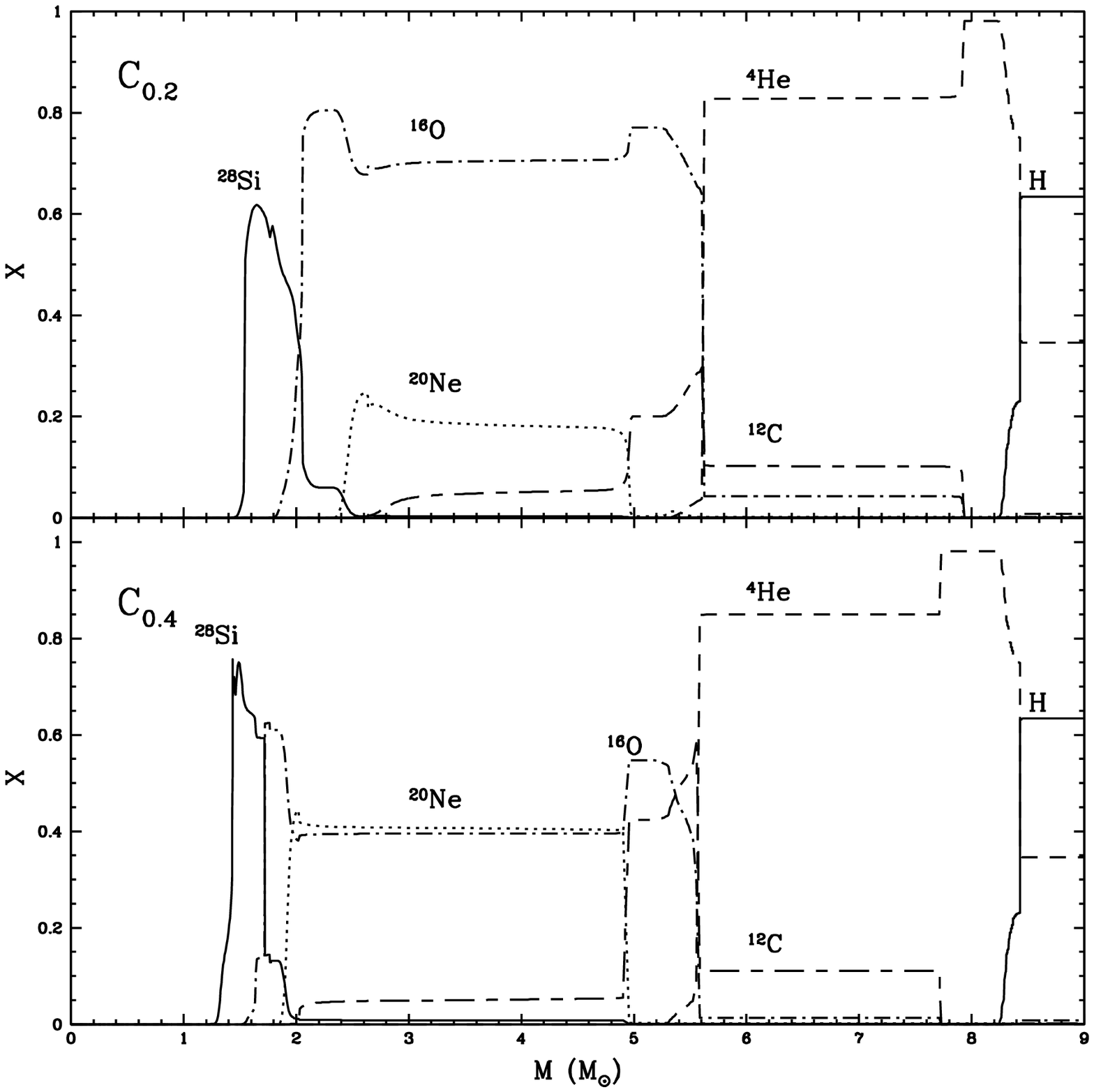}
     } 

\figcaption[f10.eps]{Structural profiles of the most abundant isotopes within the two test $\rm 25~M_\odot$ stellar models
                       at the time of the core collapse.\label{major}}

\end{figure}

\begin{figure}

\vspace*{0cm}
\mbox{ \epsfxsize=\linewidth
       \epsffile{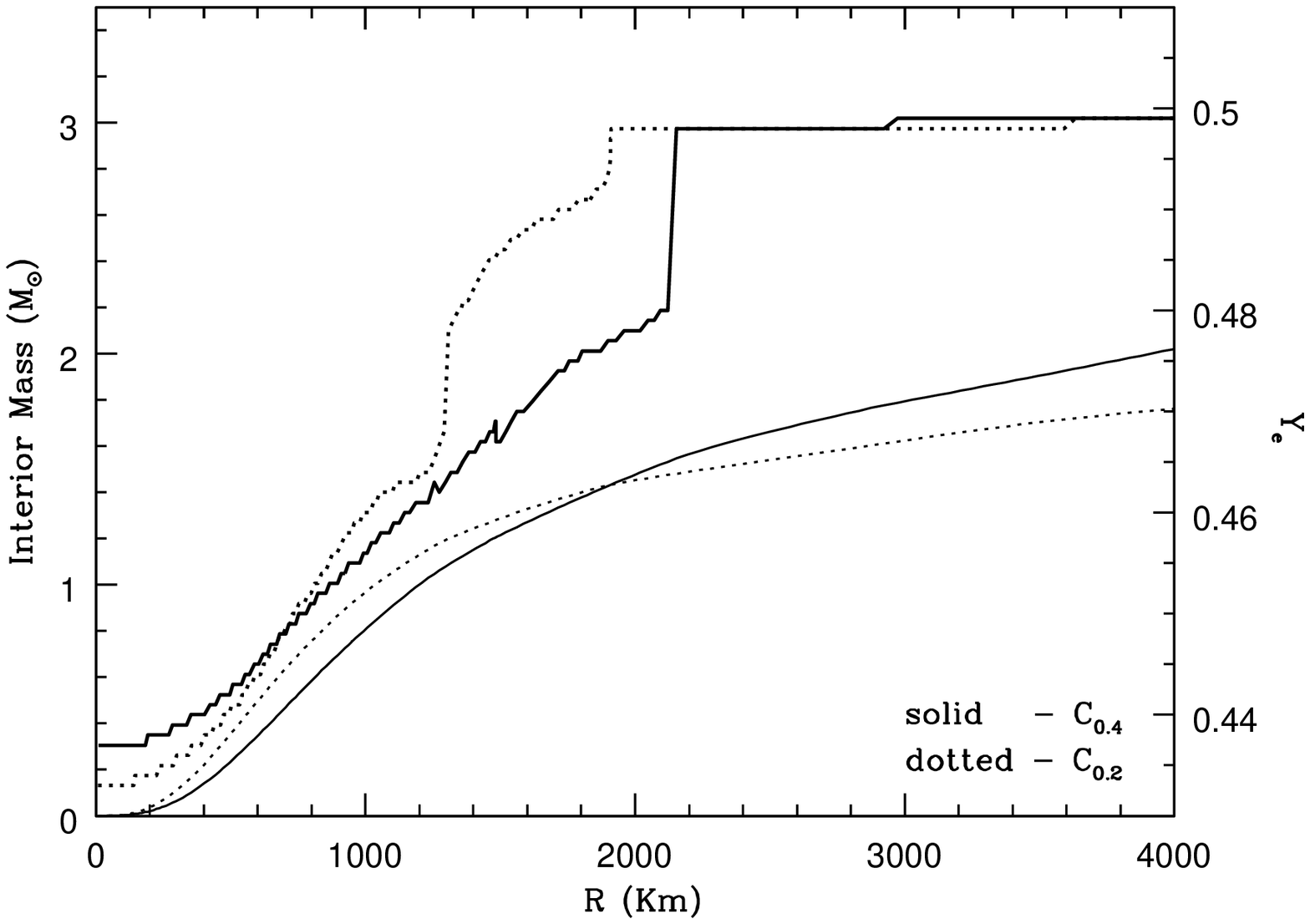}
     } 
\figcaption[f11.eps]{Comparison of the two final Mass-Radius and Ye-Radius relations.\label{M_R}}

\end{figure}

Figure \ref{tcro} shows the path followed by the two stars in the $\rm Log(t_c) \div Log(\rho_c)$ plane while Figures \ref{conv1} and \ref{conv2}
show the behavior of the convective regions as a function of time. These figures summarize the temporal 
evolution of the two stellar models up to the time of the core collapse while a snapshot of the final structure
at the time of the explosion is shown in Figures \ref{major} and \ref{M_R}: the first one shows the internal
run of the most abundant elements while the second one shows the final mass-radius relation together with the final electron mole density $\rm Y_e$.

We will not discuss in detail the properties of the various burnings and we refer the reader
to, e.g., Chieffi, Limongi and Straniero (1998) and Limongi, Straniero and Chieffi (2000) for a detailed analysis of the advanced 
burnings; here we want simply to underline how the C abundance left by the He burning, i.e. $\rm C_{ini}$,
influences the advanced burnings.

Let us firstly note that the region outside the CO core, i.e. the He and H rich layers, is not significantly
influenced by $\rm C_{ini}$ because the typical timescale on which this outer region evolves is in any case much
longer than the lifetime of all the advanced burning phases put together.

The evolution of the CO core, on the contrary, will largely depend on $\rm C_{ini}$ since both its physical and chemical
evolution will depend on the amount of fuel available in the C burning (both central and shell burnings).
Since, as it is well known, the neutrino losses become a very efficient energy sink when the central temperature
raises above $\rm \simeq 8\times 10^8~ K$, and since the formation of a convective
core requires the nuclear energy (which depends quadratically on the C abundance) to overcome the neutrino losses,
it is clear that a convective core may form in the central C burning phase only if $\rm C_{ini}$ is larger than
a threshold value. In our case a convective core forms in the 
$\rm C_{0.4}$ run while C burns in a radiative core in the $\rm C_{0.2}$ run. Once the C is exhausted in the center,
the following C shell burning occurs (in both cases)
through the formation of successive convective episodes. In spite of the very different amount of available fuel and of
the details of the shell evolution, the last
C convective shells obtained in the two cases show some conspicuous similarities: in particular the outer border of the convective shell is 
essentially insensible to $\rm C_{ini}$ because it is fixed by the location of the He shell (which is located at the same
mass coordinate in both cases) while the inner one is only mildly dependent on $\rm C_{ini}$ in the sense that the location
of the burning shell (which marks the base of the convective shell) is slightly shifted outward in the run with the lowest
initial C abundance. Roughly speaking, the size of the
convective shell reduces by almost 20\% by mass fraction by increasing the initial C abundance from 0.2 up to 0.4 dex.
The other very important similarity between the two runs is that, in spite of the very different amount of C present in the
two convective shells, both models burn almost completely the C present in the shell.
The existence of these similarities imply that the final chemical composition within 
the convective shell $\it largely$ depends on $\rm C_{ini}$. The reason is obviously that, since the C is almost completely
destroyed in both cases and since the mass size of the convective shell is similar, the 
abundances of the elements mainly produced by the C burning will directly depend on the available fuel, i.e. on $\rm C_{ini}$.

The region behind the C burning shell continues to contract (and to heat) in order to counterbalance the energy losses 
and hence to further manipulate the chemical composition (through the Ne, O and Si burnings) up to the time of the collapse.
In order to understand how the yields coming from this internal region depend on $\rm C_{ini}$ it is not necessary to discuss
in detail the various burnings beyond the C one but simply to understand how the final
mass-radius relation depends on $\rm C_{ini}$.

\begin{figure}

\vspace*{0cm}
\mbox{ \epsfxsize=\linewidth
       \epsffile{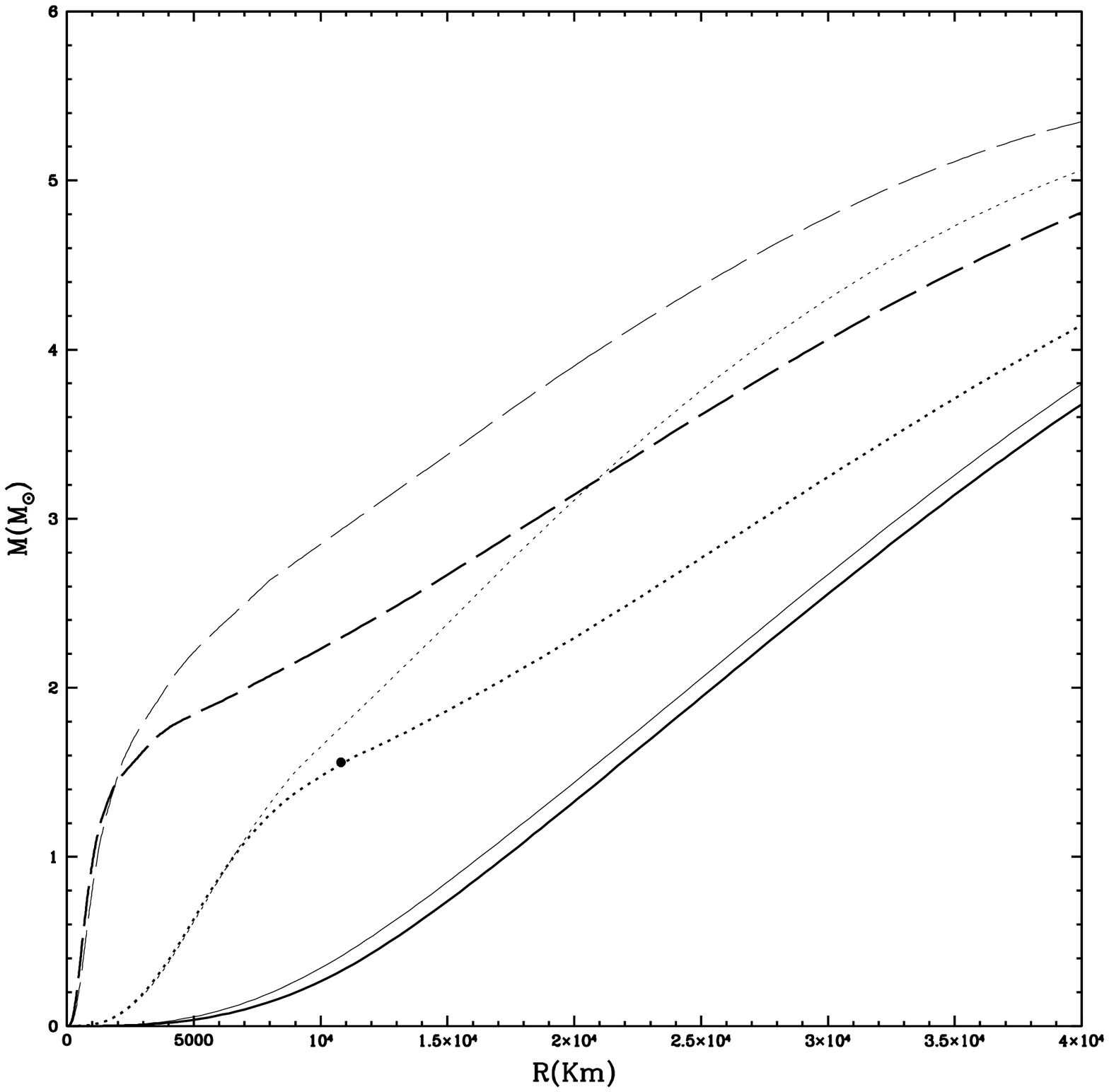}
     } 

\figcaption[f12.eps]{Mass-Radius relation at selected points along the evolution: end of the central
                       C burning (solid lines), beginning of the central Ne burning and (dotted lines) and last model (dashed lines).
                       The thin lines refer to the $\rm C_{0.2}$ case while the thick ones refer to the $\rm C_{0.4}$ run.\label{MM}}

\end{figure}

Figure \ref{MM} shows the mass-radius relation relative to the $\rm C_{\rm 0.2}$ (thin lines) and to the $\rm C_{\rm 0.4}$ (thick lines) cases
at three selected points: the solid lines mark the end of the central C burning, the dotted lines refer
to the beginning of the central Ne burning while the dashed ones refer to the last computed model.
A comparison between the two structures shows that the two models reach the end of the central
C burning with a similar M-R relation (within the first four solar masses). During the further evolutionary phases which eventually lead
to the core collapse, the very interiors of the two stars ($M\leq1.4~M_\odot$) continue to contract by maintaining a similar M-R relation,
while the more external regions reach the time of the core collapse with very different M-R relations. 
The lines showing the two M-R relations at the beginning of the central Ne burning phase reveal that most of the difference
actually start before the central Ne burning phase. In the time interval which elapses between the end of the central 
C burning and the beginning of the central Ne burning the CO core experiences a phase of contraction in which it gains from the 
gravitational field the amount of energy necessary to maintain the hydrostatic equilibrium.
In this transient phase, the energy requirement by the CO cores is partially alleviated
by the formation of one (or more) convective C shell episodes. These convective episodes stop for a while the advancing C burning front
and allow the C-burning shell to burn a "reservoir" of fuel while remaining essentially fixed in mass: such an occurrence helps in slowing
down the contraction of the region above the C burning front.
The larger amount of C available in the $\rm C_{\rm 0.4}$ case allows a more effective support of the layers 
above the C burning front and hence the formation of a M-R relation less steeper than in the other case:
the dotted lines in Figure \ref{MM} clearly show such an occurrence. Though both models will further strongly
contract up to the time of the core collapse, the differences in the M-R relations which form before the Ne
ignition remain till the final explosion.

\section{The explosive yields}

Once a pre supernova model is obtained, it is necessary to simulate in some way the explosion in order to compute the
final yields. We refer the reader to the papers by Limongi, Straniero and Chieffi (2000) and 
by Limongi, Chieffi and Straniero (in preparation) for a
comprehensive discussion of the technique we adopt to simulate the passage of a shock wave:
here it suffices to say that
the two explosions have been followed by assuming that a shock wave successfully
escapes the iron core giving a final kinetics energy of $\rm 1.2\times 10^{51}$ ergs and
that the mass cut has been arbitrarily chosen to eject $\rm 0.05~M_\odot$ of $\rm ^{56}Ni$.
In order to understand the dependence of the explosive yields on the C abundance left
by the He burning it is important to remind a few key properties of the explosion.
Once the shock wave generated by the rebounce of the core escapes the iron core, it moves through the mantle
of the star without loosing essentially any energy: hence
the peak temperature at the shock front lowers as the shock moves outward simply because it expands adiabatically and not
because it crosses progressively larger layers of the star.  This means that the peak temperature  
at the shock front is a function of its geometrical distance from the center and not of the amount of mass crossed
by the shock wave. Hence, once the energy of the shock front exiting the iron core is fixed,
it is possible to determine "a priori"
the radii at which the various peak temperatures will be reached. By reminding that at a good approximation it can 
be assumed that $\rm E=4/3 \pi R^{3} a T^{4}$,
it can be easily determined that, for an initial energy of $\rm 1.2\times10^{51}$ ergs,
a peak temperature
of $\rm 5\times 10^9~K$ is reached at r=3900 Km, a $\rm T=4\times 10^9~K$ is obtained at r=5300 Km,
a $\rm T=3.3\times 10^9~K$ is obtained at
r=6800 Km while $\rm T=1.9\times 10^9~K$ is reached at
r=14200 Km. This grid of radii corresponding to these key temperatures defines the volumes of space within which the matter 
will be exposed to, respectively, complete explosive Si burning $\rm (T_9\geq5)$, incomplete explosive Si burning
$\rm (5 \geq T_9 \geq 4)$,
explosive O burning $(\rm 4 \geq T_9 \geq 3.3)$ and C and Ne
explosive burnings $\rm (3.3 \geq T_9 \geq 1.9)$.
Apart from the C and Ne explosive burnings, all the other three explosive burnings leave a specific
(i.e. per unit mass) chemical composition which depends on the pre explosive chemical composition only through
its local degree of neutronization (which may be expressed, e.g., by means of the electron mole density $\rm Y_e$).
This means that a change in the C abundance left by the He burning does not
modify the specific yields produced by these explosive burnings (the degree of neutronization
reached by the matter in these zones is mainly determined by the conversion of $\rm ^{14}N$
- which means the initial $\rm Z_{CNO}$ - in $\rm ^{22}Ne$).
Hence $\rm C_{ini}$ influences the final yields of the elements produced by these burning only through its influence on the
final M-R relation (which means, in practice, the amount of matter located in the various key zones).
Table \ref{raggio} shows, for both runs, the amount of matter exposed to the three explosive burnings:
In accordance with the M-R relations obtained in the two cases,
the amount of matter exposed to both the explosive Oxygen burning and the incomplete explosive Si burning
is significantly larger in the $\rm C_{0.2}$ case. Only the amount of matter exposed to the complete explosive Si burning
is larger in the $\rm C_{0.4}$ case. This is, however, simply the consequence of the chosen
mass cut: in fact, the final pre-explosive structure obtained in the $\rm C_{0.2}$ run is
so compact that the required amount of $\rm ^{56}Ni$ is already almost completely synthesized by the
incomplete explosive Si burning.

\begin{deluxetable}{lcc}
\label{raggio}
\tablewidth{0pt}
\tablecaption{Mass intervals exposed to the various explosive burnings}
\tablehead{
\colhead{zone} &
\colhead{$\rm \Delta M(C_{0.2})$} &
\colhead{$\rm \Delta M(C_{0.4})$} \\
\colhead{} &
\colhead{$\rm (M_\odot)$} &
\colhead{$\rm (M_\odot)$}
}
\startdata
$Si_x$    & 0     & 0.03 \nl
$Si_{ix}$   & 0.15  & 0.09 \nl
$O_x$     & 0.22  & 0.13 \nl
$(C\&Ne)_x$  & 0.82  & 0.68 \nl
\enddata
\end{deluxetable}                                                                               

Keeping in mind these properties of the explosion we can now turn to the analysis of the dependence of
the explosive yields on $\rm C_{ini}$.

\begin{deluxetable}{llllll}
\label{tab_yields}

\tabletypesize{\scriptsize} 
\tablewidth{0pt}

\tablecaption{\small \bf Isotopic yields $2.5\times 10^4 ~ s$ after the rebounce.}

\tablehead{
\colhead{} &
\colhead{$\rm C_{0.2}$} &
\colhead{$\rm C_{0.4}$} &
\colhead{} &
\colhead{$\rm C_{0.2}$} &
\colhead{$\rm C_{0.4}$} \\
\colhead{} &                        
\colhead{$\rm (M_\odot)$} &         
\colhead{$\rm (M_\odot)$}&          
\colhead{} &                        
\colhead{$\rm (M_\odot)$} &         
\colhead{$\rm (M_\odot)$}          
}
\startdata 
$\rm ^{} H     $     & $     1.04\times 10^{+1}  $    &$  1.04\times 10^{+1}  $   &  $\rm ^{39} K   $     & $     2.07\times 10^{-4}  $    &$  1.58\times 10^{-4}  $      \\
$\rm ^{2}  H   $     & $     2.20\times 10^{-16} $    &$  2.19\times 10^{-16} $   &  $\rm ^{40} K   $     & $     3.04\times 10^{-6}  $    &$  4.14\times 10^{-6}  $     \\
$\rm ^{3}  He  $     & $     2.79\times 10^{-4}  $    &$  2.79\times 10^{-4}  $   &  $\rm ^{41} K   $     & $     2.18\times 10^{-5}  $    &$  1.78\times 10^{-5}  $     \\
$\rm ^{4}  He  $     & $     8.02\times 10^{ 0}  $    &$  8.13\times 10^{ 0}  $   &  $\rm ^{40} Ca  $     & $     1.30\times 10^{-2}  $    &$  6.83\times 10^{-2}  $      \\
$\rm ^{6}  Li  $     & $     1.33\times 10^{-9}  $    &$  1.33\times 10^{-9}  $   &  $\rm ^{42} Ca  $     & $     8.36\times 10^{-5}  $    &$  5.16\times 10^{-5}  $      \\
$\rm ^{7}  Li  $     & $     6.74\times 10^{-11} $    &$  6.79\times 10^{-11} $   &  $\rm ^{43} Ca  $     & $     5.12\times 10^{-6}  $    &$  5.94\times 10^{-6}  $       \\
$\rm ^{9}  Be  $     & $     3.39\times 10^{-10} $    &$  3.40\times 10^{-10} $   &  $\rm ^{44} Ca  $     & $     4.60\times 10^{-5}  $    &$  5.76\times 10^{-5}  $       \\
$\rm ^{10} B   $     & $     2.25\times 10^{-9}  $    &$  2.25\times 10^{-9}  $   &  $\rm ^{46} Ca  $     & $     1.74\times 10^{-6}  $    &$  2.77\times 10^{-7}  $       \\
$\rm ^{11} B   $     & $     2.07\times 10^{-8}  $    &$  2.08\times 10^{-8}  $   &  $\rm ^{48} Ca  $     & $     3.04\times 10^{-6}  $    &$  3.11\times 10^{-6}  $      \\ 
$\rm ^{12} C   $     & $     5.04\times 10^{-1}  $    &$  6.83\times 10^{-1}  $   &  $\rm ^{45} Sc  $     & $     4.51\times 10^{-6}  $    &$  3.31\times 10^{-6}  $      \\ 
$\rm ^{13} C   $     & $     2.22\times 10^{-3}  $    &$  2.24\times 10^{-3}  $   &  $\rm ^{46} Ti  $     & $     3.12\times 10^{-5}  $    &$  2.13\times 10^{-5}  $      \\ 
$\rm ^{14} N   $     & $     7.87\times 10^{-2}  $    &$  8.11\times 10^{-2}  $   &  $\rm ^{47} Ti  $     & $     6.48\times 10^{-6}  $    &$  6.83\times 10^{-6}  $      \\ 
$\rm ^{15} N   $     & $     2.70\times 10^{-5}  $    &$  2.71\times 10^{-5}  $   &  $\rm ^{48} Ti  $     & $     1.65\times 10^{-4}  $    &$  1.18\times 10^{-4}  $      \\ 
$\rm ^{16} O   $     & $     2.39\times 10^{0}   $    &$  1.71\times 10^{0}   $   &  $\rm ^{49} Ti  $     & $     1.74\times 10^{-5}  $    &$  1.43\times 10^{-5}  $      \\ 
$\rm ^{17} O   $     & $     1.29\times 10^{-4}  $    &$  1.30\times 10^{-4}  $   &  $\rm ^{50} Ti  $     & $     1.28\times 10^{-5}  $    &$  1.45\times 10^{-5}  $       \\
$\rm ^{18} O   $     & $     5.04\times 10^{-4}  $    &$  2.56\times 10^{-4}  $   &  $\rm ^{50} V  $     & $     1.48\times 10^{-7}  $    &$  1.52\times 10^{-7}  $        \\
$\rm ^{19} F   $     & $     1.27\times 10^{-5}  $    &$  1.09\times 10^{-5}  $   &  $\rm ^{51} V  $     & $     3.04\times 10^{-5}  $    &$  2.23\times 10^{-5}  $        \\
$\rm ^{20} Ne  $     & $     3.53\times 10^{-1}  $    &$  1.02\times 10^{0}   $   &  $\rm ^{50} Cr  $     & $     1.79\times 10^{-4}  $    &$  1.25\times 10^{-4}  $     \\     
$\rm ^{21} Ne  $     & $     2.18\times 10^{-3}  $    &$  1.83\times 10^{-3}  $   &  $\rm ^{52} Cr  $     & $     2.48\times 10^{-3}  $    &$  1.44\times 10^{-3}  $     \\ 
$\rm ^{22} Ne  $     & $     5.93\times 10^{-2}  $    &$  5.30\times 10^{-2}  $   &  $\rm ^{53} Cr  $     & $     2.71\times 10^{-4}  $    &$  1.78\times 10^{-4}  $     \\ 
$\rm ^{23} Na  $     & $     1.60\times 10^{-2}  $    &$  3.23\times 10^{-2}  $   &  $\rm ^{54} Cr  $     & $     3.24\times 10^{-5}  $    &$  3.53\times 10^{-5}  $     \\ 
$\rm ^{24} Mg  $     & $     8.21\times 10^{-2}  $    &$  2.99\times 10^{-1}  $   &  $\rm ^{55} Mn  $     & $     1.22\times 10^{-3}  $    &$  8.79\times 10^{-3}  $     \\ 
$\rm ^{25} Mg  $     & $     2.40\times 10^{-2}  $    &$  3.41\times 10^{-2}  $   &  $\rm ^{54} Fe  $     & $     1.41\times 10^{-2}  $    &$  1.01\times 10^{-2}  $     \\ 
$\rm ^{26} Mg  $     & $     1.86\times 10^{-2}  $    &$  2.61\times 10^{-2}  $   &  $\rm ^{56} Fe  $     & $     7.36\times 10^{-2}  $    &$  7.38\times 10^{-2}  $      \\
$\rm ^{27} Al  $     & $     1.26\times 10^{-2}  $    &$  3.20\times 10^{-2}  $   &  $\rm ^{57} Fe  $     & $     1.74\times 10^{-3}  $    &$  2.28\times 10^{-3}  $      \\                
$\rm ^{28} Si  $     & $     2.09\times 10^{-1}  $    &$  1.87\times 10^{-1}  $   &  $\rm ^{58} Fe  $     & $     1.06\times 10^{-3}  $    &$  1.15\times 10^{-3}  $     \\       
$\rm ^{29} Si  $     & $     6.06\times 10^{-3}  $    &$  7.65\times 10^{-3}  $   &  $\rm ^{59} Co  $     & $     4.49\times 10^{-4}  $    &$  6.48\times 10^{-4}  $     \\                 
$\rm ^{30} Si  $     & $     6.69\times 10^{-3}  $    &$  7.57\times 10^{-3}  $   &  $\rm ^{58} Ni  $     & $     2.17\times 10^{-3}  $    &$  3.32\times 10^{-3}  $      \\ 
$\rm ^{31} P   $     & $     1.78\times 10^{-3}  $    &$  2.20\times 10^{-3}  $   &  $\rm ^{60} Ni  $     & $     9.36\times 10^{-4}  $    &$  1.02\times 10^{-3}  $      \\                                                 
$\rm ^{32} S   $     & $     1.06\times 10^{-1}  $    &$  7.00\times 10^{-2}  $   &  $\rm ^{61} Ni  $     & $     2.04\times 10^{-4}  $    &$  2.30\times 10^{-4}  $       \\                                                               
$\rm ^{33} S   $     & $     5.72\times 10^{-4}  $    &$  6.05\times 10^{-4}  $   &  $\rm ^{62} Ni  $     & $     5.50\times 10^{-4}  $    &$  6.62\times 10^{-4}  $       \\                                                                                 
$\rm ^{34} S   $     & $     6.74\times 10^{-3}  $    &$  7.06\times 10^{-3}  $   &  $\rm ^{64} Ni  $     & $     4.96\times 10^{-4}  $    &$  4.83\times 10^{-4}  $       \\                                                                                 
$\rm ^{36} S   $     & $     2.83\times 10^{-5}  $    &$  2.99\times 10^{-5}  $   &  $\rm ^{63} Cu  $     & $     2.85\times 10^{-4}  $    &$  3.04\times 10^{-4}  $      \\                                                                                  
$\rm ^{35} Cl  $     & $     2.45\times 10^{-4}  $    &$  2.43\times 10^{-4}  $   &  $\rm ^{65} Cu  $     & $     1.20\times 10^{-4}  $    &$  1.40\times 10^{-4}  $      \\ 
$\rm ^{37} Cl  $     & $     1.92\times 10^{-4}  $    &$  2.10\times 10^{-4}  $   &  $\rm ^{64} Zn  $     & $     5.61\times 10^{-5}  $    &$  1.14\times 10^{-4}  $      \\ 
$\rm ^{36} Ar  $     & $     1.64\times 10^{-2}  $    &$  9.40\times 10^{-3}  $   &  $\rm ^{66} Zn  $     & $     1.71\times 10^{-4}  $    &$  2.65\times 10^{-4}  $      \\ 
$\rm ^{38} Ar  $     & $     3.03\times 10^{-3}  $    &$  2.25\times 10^{-3}  $   &  $\rm ^{67} Zn  $     & $     2.98\times 10^{-5}  $    &$  5.13\times 10^{-5}  $      \\ 
$\rm ^{40} Ar  $     & $     6.37\times 10^{-6}  $    &$  6.06\times 10^{-6}  $   &  $\rm ^{68} Zn  $     & $     4.97\times 10^{-4}  $    &$  7.04\times 10^{-4}  $      
 
\enddata          
       
\end{deluxetable}

\begin{deluxetable}{lll}
\label{tab_ele}
\tabletypesize{\scriptsize} 

\tablewidth{0pt}
\tablecaption{Elemental yields fully decayed.}
\tablehead{
\colhead{} &
\colhead{$\rm C_{0.2}$} &
\colhead{$\rm C_{0.4}$} \\
\colhead{} &
\colhead{$\rm (M_\odot)$} &
\colhead{$\rm (M_\odot)$}
}
\startdata

H       & $     1.04\times 10^{+01} $   & $      1.04\times 10^{+01 } $      \\
He      & $     8.14\times 10^{+00} $   & $      8.02\times 10^{+00 } $      \\
C       & $     6.85\times 10^{-01} $   & $      5.06\times 10^{-01 } $      \\
N       & $     8.11\times 10^{-02} $   & $      7.88\times 10^{-02 } $      \\
O       & $     1.71\times 10^{+00} $    &$       2.39\times 10^{+00} $      \\
F       & $     1.09\times 10^{-05} $    &$       1.27\times 10^{-05} $       \\
Ne      & $     1.08\times 10^{+00} $    &$       4.14\times 10^{+01} $       \\
Na      & $     3.23\times 10^{-02} $    &$       1.60\times 10^{-02} $       \\
Mg      & $     3.60\times 10^{-01} $    &$       1.25\times 10^{-01} $      \\ 
Al      & $     3.20\times 10^{-02} $    &$       1.26\times 10^{-02} $      \\ 
Si      & $     2.02\times 10^{-01} $    &$       2.22\times 10^{-01} $      \\ 
P       & $     2.20\times 10^{-03} $    &$       1.78\times 10^{-03} $      \\ 
S       & $     7.77\times 10^{-02} $    &$       1.13\times 10^{-01} $      \\ 
Cl      & $     4.53\times 10^{-04} $    &$       4.36\times 10^{-04} $      \\ 
Ar      & $     1.17\times 10^{-02} $    &$       1.94\times 10^{-02} $       \\
K       & $     1.80\times 10^{-04} $    &$       2.32\times 10^{-04} $       \\
Ca      & $     6.94\times 10^{-03} $    &$       1.31\times 10^{-02} $       \\
Sc      & $     3.31\times 10^{-06} $    &$       4.51\times 10^{-06} $     \\     
Ti      & $     1.75\times 10^{-04} $    &$       2.33\times 10^{-04} $     \\ 
V       & $     2.25\times 10^{-05} $    &$       3.05\times 10^{-05} $     \\ 
Cr      & $     1.78\times 10^{-03} $    &$       2.96\times 10^{-03} $     \\ 
Mn      & $     8.79\times 10^{-04} $    &$       1.22\times 10^{-03} $     \\ 
Fe      & $     8.73\times 10^{-02} $    &$       9.05\times 10^{-02} $     \\ 
Co      & $     6.48\times 10^{-04} $    &$       4.49\times 10^{-04} $      \\
Ni      & $     5.71\times 10^{-03} $    &$       4.36\times 10^{-03} $      \\ 
Cu      & $     4.45\times 10^{-04} $    &$       4.05\times 10^{-04} $      \\                
Zn      & $     1.13\times 10^{-03} $    &$       7.54\times 10^{-04} $       \\                   

\enddata
\end{deluxetable}

\begin{figure}

\vspace*{0cm}
\mbox{ \epsfxsize=\linewidth
       \epsffile{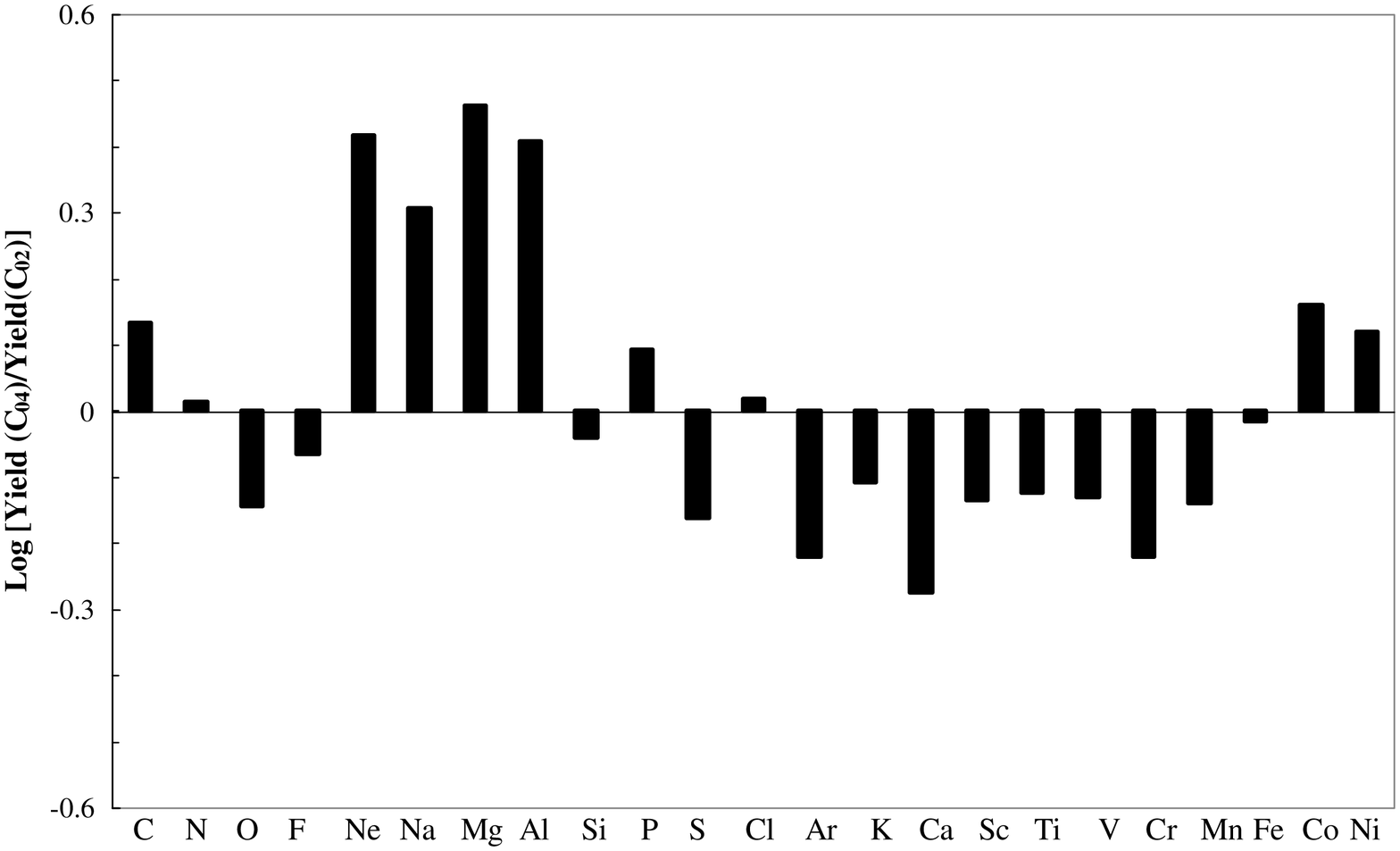}
     } 
\figcaption[f13.eps]{The logarithmic ratio between the Yields produced by the $\rm C_{0.4}$ run and those produced by the $\rm C_{0.2}$
                       run.\label{isto_ele}}

\end{figure}

Table \ref{tab_yields} shows the isotopic yields ($\rm 2.5\times 10^4$ s after the rebounce)
produced in the two cases while table \ref{tab_ele} and Figure \ref{isto_ele} show a comparison
between the elemental (fully decayed) yields.
In the following
we will focus our attention only on the elemental yields. This will be equivalent, in general, to speak about the most abundant isotope:
of course if more than one isotope is important to describe the behavior of an element we will address all the important ones. 
A proper discussion of Figure \ref{isto_ele} requires the knowledge of the production site of each element.
Schematically, four main groups of elements may be identified (the isotopes within the brackets at the right side of each element
show the main isotopes, at least in these runs, which determine the final elemental yields): 
the first one, which includes Ne ($\rm ^{20}Ne$), Na ($\rm ^{23}Na$), Mg ($\rm ^{24}Mg$), 
Al ($\rm ^{27}Al$), P ($\rm ^{31}P$),
Cl ($\rm ^{35}Cl$ and $\rm ^{37}Cl$) and Sc ($\rm ^{45}Sc$ and $\rm ^{45}Ca$), is produced in the C convective shell;
the second one (i.e. the "golden" group, see Limongi, Chieffi and Straniero $\it in ~preparation$) is produced by both
the incomplete explosive Si burning and the explosive O burning and includes Si ($\rm ^{28}Si$), S ($\rm ^{32}S$),
Ar ($\rm ^{36}Ar)$, Ca  ($\rm ^{40}Ca$) plus K ($\rm ^{39}K$)
which is synthesized only by the explosive O burning); the third one is produced only by the explosive 
incomplete Si burning and includes Ti ($\rm ^{48}Cr$), V ($\rm ^{51}Cr$), Cr ($\rm ^{52}Cr$, $\rm ^{52}Mn$ and $\rm ^{52}Fe$) and Mn ($\rm ^{55}Mn$,
$\rm ^{55}Fe$ and $\rm ^{55}Co$); the last one is produced by the complete explosive Si burning and includes 
Fe ($\rm ^{56}Ni$, $\rm ^{56}Fe$ and $\rm ^{54}Fe$), Co ($\rm ^{59}Co$), and Ni ($\rm ^{58}Ni$, $\rm ^{60}Ni$, $\rm ^{61}Ni$, 
$\rm ^{62}Ni$ and $\rm ^{64}Ni$) 
(We can't discuss Cu and Zn because they are just the upper end of our network).
Iron is actually produced also as $\rm ^{56}Ni$ by the incomplete explosive Si burning.
The light elements H to F will be discussed separately.
Note that all the following discussion strictly holds only for the $\rm 25~M_\odot$ even if it probably
may be considered more or less valid in general.

\begin{figure}

\vspace*{0cm}
\mbox{ \epsfxsize=\linewidth
       \epsffile{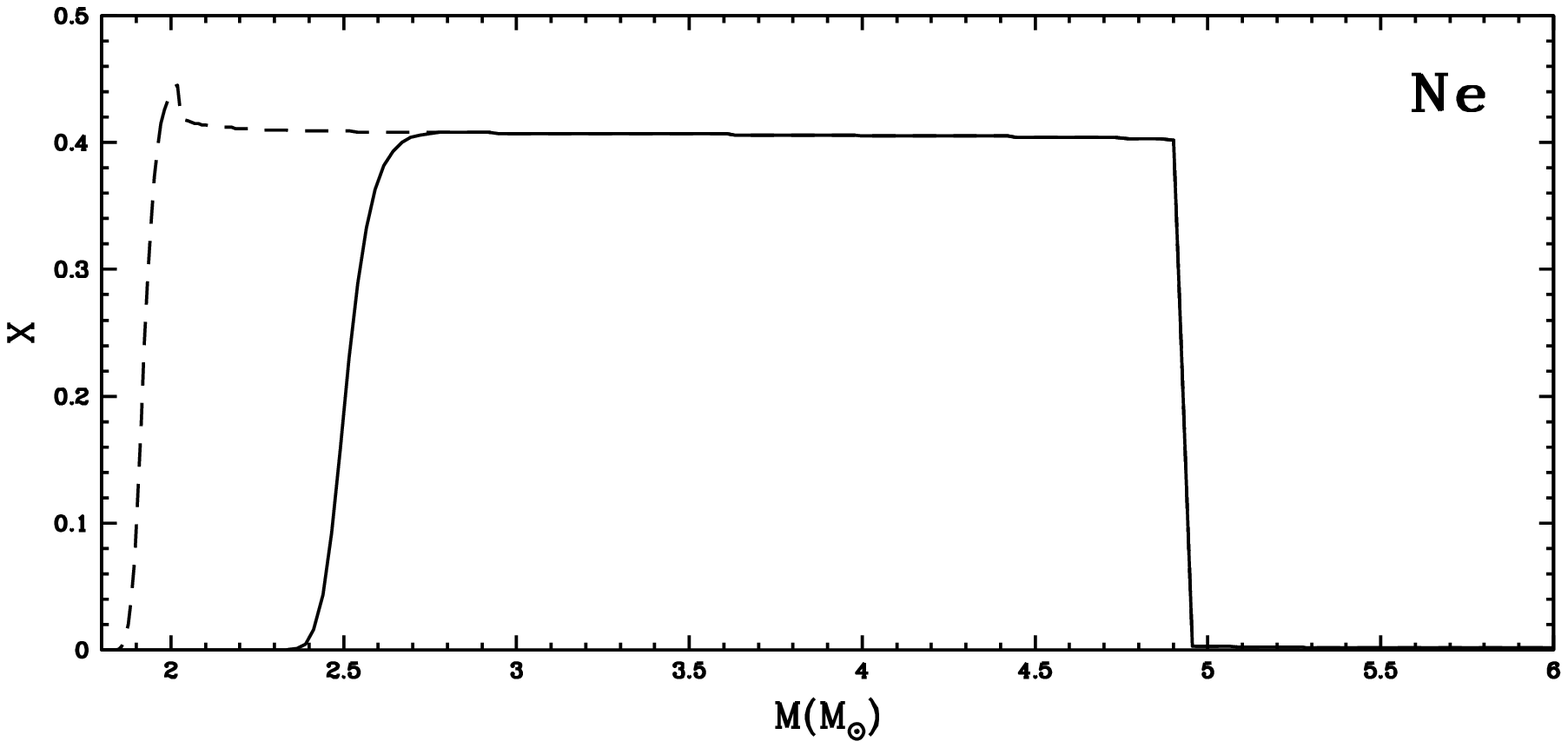}
     } 

\figcaption[f14.eps]{Neon profile ($\rm C_{0.4}$ case) within the star at the time of the explosion (dashed line) and after the passage of the shock wave
                       (solid line).\label{neon}}

\end{figure}

Four out of the seven elements pertaining to the first group, namely Ne, Na, Mg and Al, present a very similar behavior:
they are produced in the C convective shell and are partially destroyed by the explosion. Hence their final yields largely depend on the 
pre explosive ones. Figure \ref{neon} shows,
as a typical example, the Ne profile at the time of the core collapse (dashed line) and after the passage of the shock wave (solid line).
All four elements show a similar dependence on $\rm C_{ini}$, in the sense that they scale more or less uniformly (and directly) 
with the C abundance left by 
the He burning (see Figure \ref{isto_ele}). However this occurrence is somewhat accidental. Ne and Mg are direct products of C burning and hence it is
quite reasonable that they scale similarly with $\rm C_{ini}$. Na, on the contrary, though it is a primary product of C burning, 
settles rapidly at its equilibrium value between production and destruction: this equilibrium value is almost 
independent on the initial C abundance  while it largely depends on the temperature in the sense
that its abundance increases as the temperature decreases. Since the temperature at the base of the C convective shell 
scales inversely with $\rm C_{ini}$ it follows that the Na yield increases as $\rm C_{ini}$ increases (hence behaving similarly to Ne and Mg). 
What it is quite accidental is that it scales even quantitatively in a way similar to that of Ne and Mg.
Al shows an even different behavior: it has a secondary origin (i.e. it descends mainly from the initial
abundance of CNO) and it is formed mainly by the sequence 
$\rm ^{22}Ne(\alpha,n)^{25}Mg(n,\gamma)^{26}Mg(p,\gamma)^{27}Al$
plus the additional (primary) sequence 
$\rm ^{23}Na(\alpha,p)^{26}Mg(p,\gamma)^{27}Al$.
Since the first (dominant) sequence originates from the $\rm ^{22}Ne$ (plus $\rm ^{25}Mg$ and $\rm ^{26}Mg$), 
one could expect the Al yield to be controlled first
of all by the amount of  $\rm ^{22}Ne+^{25}Mg+^{26}Mg$, i.e. the seed nuclei. This is not the case because the starting reaction of this sequence requires
the presence of $\alpha$ particles (which in this conditions come directly from the $\rm ^{12}C+^{12}C$ reactions); since, even for the largest initial
C abundances, not all the $\rm ^{22}Ne+^{25}Mg+^{26}Mg$ is fully converted in Al, it happens that the final Al production scales with the $\rm \alpha$'s
available, which means that it scales with the initial C abundance. Note that, as the initial metallicity reduces, the second (primary)
channel becomes more important so that the final Al abundance always depends on $\rm C_{ini}$ and hence it behaves as a primary element. 
Let us now turn to the other three elements pertaining to the first group: i.e. P, Cl and Sc. Each of them has a specific history and hence
they will be discussed individually.

\begin{figure}

\vspace*{0cm}
\mbox{ \epsfxsize=\linewidth
       \epsffile{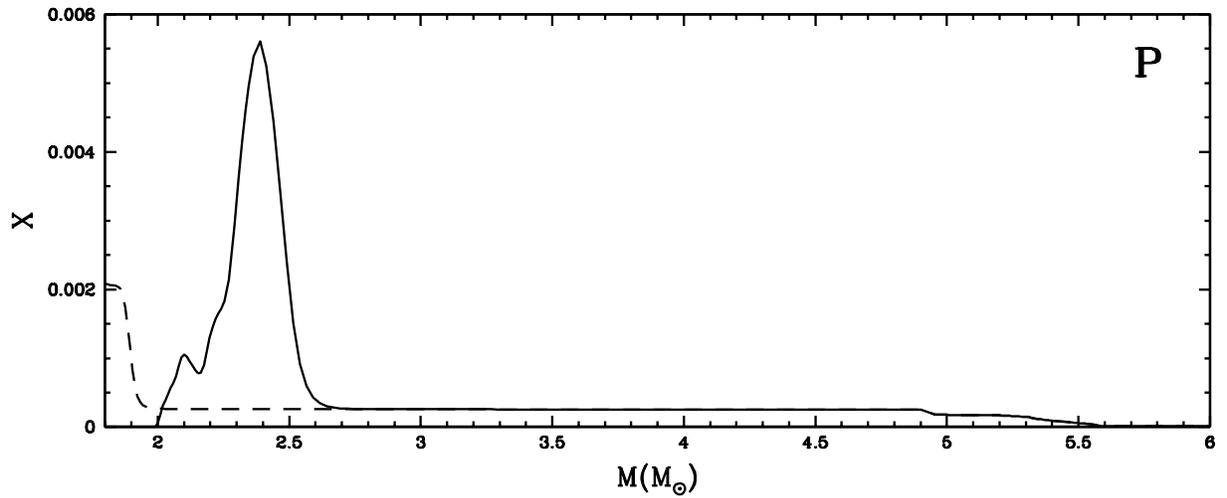}
     } 
\figcaption[f15.eps]{Phosphorus profile ($\rm C_{0.4}$ case) within the star at the time of the explosion (dashed line) and after the passage of the shock wave
                       (solid line).\label{phos}}

\end{figure}

P is produced by the Ne burning and its site of hydrostatic production is deep enough that it is completely
destroyed by the passage of the shock wave: the P which will be ejected by the explosion is only the one directly produced
by the shock itself when passing through the Ne rich layers: its typical final profile is shown in Figure \ref{phos}.
It is possible to identify two main chains of processes which lead to the production of P, the first one, primary, starts directly from
the $\rm ^{20}Ne$ and leads through several sub channels to P, and a second one which starts from $\rm ^{25}Mg$ and hence it is a typical secondary sequence.
The not negligible presence of a true secondary component which does not depend on $\rm C_{ini}$ leads to a rather mild global dependence
of P on $\rm C_{ini}$.

Cl is a complex element since both stable isotopes, $\rm ^{35}Cl$ and $\rm ^{37}Cl$,
contribute significantly to the Cl elemental yield. $\rm ^{35}Cl$ is similar to P because it is 
produced by the explosive burning 
while $\rm ^{37}Cl$ is produced in the central He burning phase and it is then preserved down to the 
base of the C convective shell
up to the time of the core collapse. The passage of the shock wave partially destroys the $\rm ^{37}Cl$,
in a manner similar to Ne. Since none of the two isotopes descends directly from $\rm C_{ini}$, the final
abundance of this element is practically unaffected by a change in $\rm C_{ini}$.

Sc, the last element pertaining to this group, scales inversely with $\rm C_{ini}$. Its production
depends on the abundance of $\rm ^{45}Sc$ itself and $\rm ^{45}Ca$ which decays in $\rm ^{45}Sc$.
Such a scaling may be explained by noting that
$\rm ^{45}Ca$ is a branching point in which two concurrent process compete, i.e. the $\beta$ decay and the neutron capture,
and hence that the final fate of the $\rm ^{45}Ca$ depends on the neutron density; this last quantity scales 
directly with the C abundance simply because the same total amount of neutrons is released on different timescales.
In the $\rm C_{0.2}$ case the neutron flux is large enough that most of the matter coming from $\rm ^{44}Ca$ goes to 
$\rm ^{45}Ca$ first and to $\rm ^{46}Ca$ later:
the abundance of $\rm ^{45}Ca$ is in this case high and determined by the equilibrium condition between 
$\rm ^{44}Ca(n,\gamma)$ and $\rm ^{45}Ca(n,\gamma)$. 
In the $\rm C_{0.4}$ case, on the contrary, the low neutron flux allows the $\rm ^{45}Ca$ to decay in $\rm ^{45}Sc$, so that
the final abundance 
of $\rm ^{45}Sc$ is settled by the competition between $\rm ^{44}Ca(n,\gamma)$ and $\rm ^{45}Sc(n,\gamma)$.

Four out of the five elements pertaining to the Golden group, namely Si, S, Ar and Ca, have a similar history.
Since they are synthesized by both the incomplete explosive Si burning and the explosive O burning,
their yields do not depend on the pre explosive composition
(apart from the neutron excess which is however the same in the two runs) but only on the amount of mass exposed to these burnings. Since,
as we have shown above, the
lower the C abundance the larger is the amount of mass exposed to these explosive burnings, the final yields of these elements scale
inversely with $\rm  C_{ini}$. The relative abundances of these elements change mildly, but systematically, by a
change in $\rm C_{ini}$ so that the four elements (Si, S, Ar and Ca) are progressively more overproduced going from
the $\rm C_{0.4}$ to the $\rm  C_{0.2}$ case.

The yields of the elements produced by the incomplete explosive Si burning depend on the amount of mass exposed to 
a peak temperature in the range 4-3 billions degree and hence they
scale inversely with $\rm C_{ini}$. The opposite behavior of Ni and Co can be easily understood by noting that
they are mainly produced by the complete explosive Si burning and that essentially no matter coming from this region
is ejected in the $\rm  C_{0.2}$ case (this is obviously true only for this specific choice of the mass cut).

Let us now address also the dependence of the yields of the 
three long-lived radioactive isotopes, $\rm ^{26}Al$, $\rm ^{60}Fe$ and $\rm ^{44}Ti$ on $\rm C_{ini}$ 
because of their importance as gamma ray emitters. These three isotopes
are produced by the explosion itself, since their pre explosive abundance is either negligible or destroyed
by the passage of the shock wave.
The first two, i.e. $\rm ^{26}Al$ and $\rm ^{60}Fe$, are produced in the region where the peak temperature 
of the shocks reaches a value
of the order of $\rm 2\times10^9~K$. The amount of $\rm ^{26}Al$ produced depends on the abundances of $\rm ^{25}Mg$, 
$\rm ^{23}Na$ and $\rm ^{20}Ne$ (or $\rm ^{12}C$)
because the $\rm \alpha$ particles produced by the Ne (and/or C) burning are captured by $\rm ^{23}Na$ which releases the protons
which are then partially captured by $\rm ^{25}Mg$ to produce $\rm ^{26}Al$.
Hence the yield of this isotope scales directly with $\rm C_{ini}$ and varies from
$\rm 3.28\times 10^{-5}$ (for $\rm C_{0.4}$) to $2.67\times 10^{-5}$ (for $\rm C_{0.2}$).
The production of $\rm ^{60}Fe$ requires a double neutron capture on the stable isotope $\rm ^{58}Fe$.
Neutrons are mainly produced by the capture of $\rm \alpha$ particles on $\rm ^{22}Ne$ and hence it is the abundance
of this isotope at the time of the core collapse (in the region where the peak temperature is
of the order of $\rm 2\times10^9~K$) which controls the yield of the $\rm ^{60}Fe$.
Since the final abundance of $\rm ^{22}Ne$
scales inversely with $\rm C_{ini}$, also the final $\rm ^{60}Fe$ abundance will follow the same trend.
In particular we predict an $\rm ^{60}Fe$ yield equal to $\rm 4.94\times 10^{-6}$ ($\rm C_{0.4}$ case)
and to $\rm 2.92\times 10^{-5}$ ($\rm C_{0.2}$ case).
$\rm ^{44}Ti$ is produced by the complete explosive Si burning and hence its yield will depend, among the other things,
on the mass cut and on the degree of freeze-out experienced by the most internal layers of the ejecta. Under the (arbitrary)
assumption that in both runs $\rm 0.05~M_\odot$ of $\rm ^{56}Ni$ are ejected, we obtain that the 
yield of this isotope scales inversely with $\rm C_{ini}$, and in particular that it reduces from $\rm 1.1\times 10^{-5}$ 
to $\rm 3.4\times 10^{-6}$.

The light elements C and O are only marginally affected by the explosion and hence they reflect 
essentially their pre explosive abundances: it goes without saying that while the C yield scales 
directly with $\rm C_{ini}$, the O yield scales inversely with $\rm C_{ini}$.
N is produced in the H burning and hence it is not affected by $\rm C_{ini}$.
Fluorine is synthesized in the He burning shell by the chain
$\rm ^{14}N(\alpha,\gamma)^{18}F(\beta)^{18}O(p,\alpha)^{15}N(\alpha,\gamma)^{19}F$ and scales
inversely with $\rm C_{ini}$ because of both the mass size of the He shell and the $\rm ^{18}O$ abundance
scale inversely with $\rm C_{ini}$.

\section{Final discussion and conclusions}

\begin{figure}

\vspace*{0cm}
\mbox{ \epsfxsize=\linewidth
       \epsffile{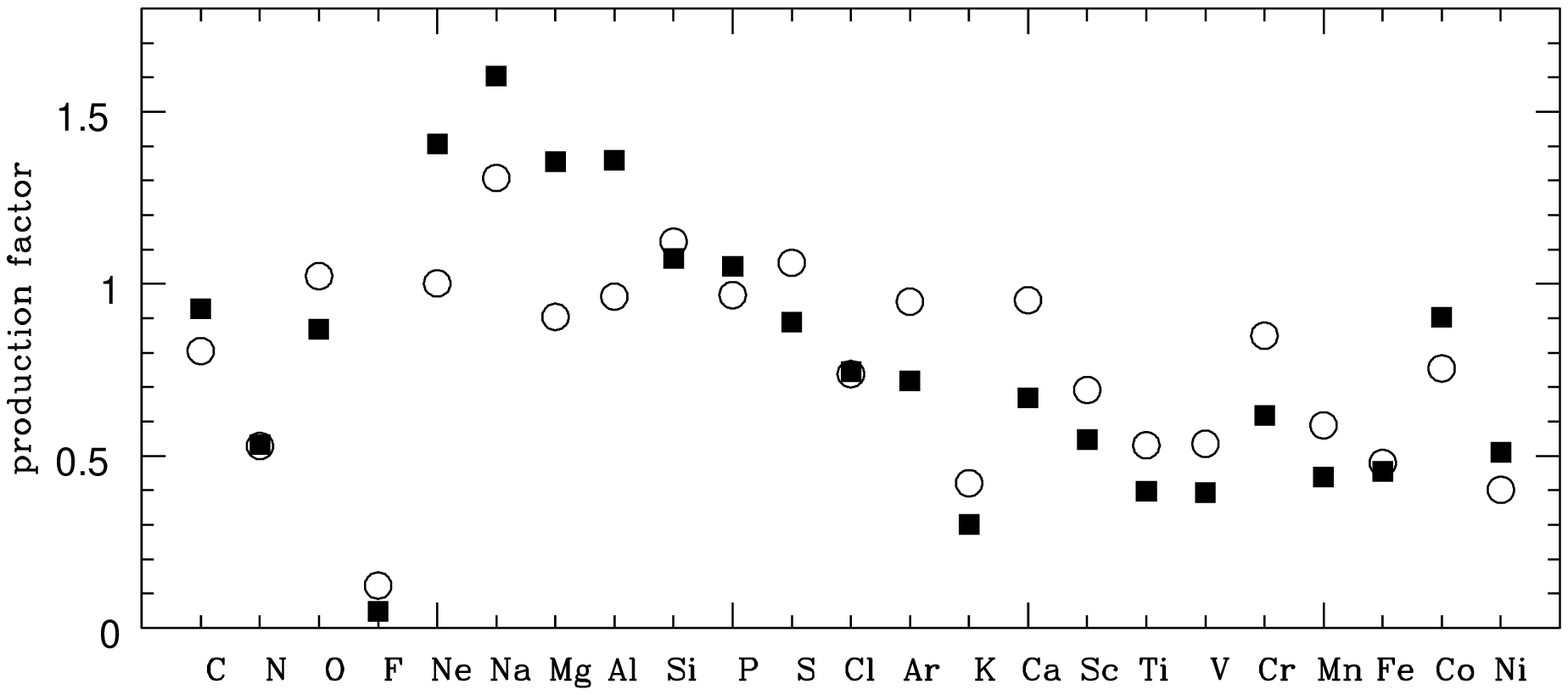}
     } 

\figcaption[f16.eps]{Comparison between the production factors obtained in the $\rm C_{0.4}$ case (filled squares) and those
                       produced in the $\rm C_{0.2}$ case (open dots).\label{over}}

\end{figure}

In the previous section we have shown that the elements between Ne and Ni may be divided in groups of nuclei
which behave more or less similarly with respect to a change in $\rm C_{ini}$. Roughly speaking we can say 
that while the elements produced in the C convective shell scale directly with $\rm C_{ini}$ (the larger 
the C abundance left by the He burning the larger the final abundance of these elements),
most of the elements synthesized by the explosive burnings scale inversely with $\rm C_{ini}$
because of the steeper M-R relation. Hence a comparison between the solar chemical composition 
and the yields obtained with the two values of $\rm C_{ini}$ could help
to constrain the real abundance of C left by the He burning. However, in order to obtain a robust comparison it would be
necessary to integrate at least over a stellar generation extending between $\rm 13$ and $\rm 30~M_\odot$; at present
we do not have such an extended set of computations for two different values of $\rm C_{ini}$. Nonetheless we think 
that it is in any case interesting to show a simple comparison between our data and the solar chemical distribution 
because, as it has already been noted several times (e.g. \citet{WW82}), the star which influences more pronouncedly
the chemical composition of the ejecta of a generation of stars is of the order of the $25~M_\odot$ if the adopted IMF
is the Salpeter one. Figure \ref{over} shows the production factors of all the
elements discussed above for the two values of $\rm C_{ini}$: the filled squares refer to the $\rm C_{0.4}$ case while the open dots 
refer to the $\rm C_{0.2}$ one.

By the way, the production factor is defined as the ratio between the amount of mass (in solar masses)
ejected as a given element and the amount of mass (in solar masses) the same elements would have if all
the ejecta would have a solar chemical composition. It goes without saying that
all the elements sharing a similar production factor maintain scaled solar
relative proportions. 

In the $\rm C_{0.4}$ case it can be seen that, even if slightly, C is overproduced with respect to Oxygen;
this would imply that, at the very least, there would not be room for C production by other kinds of stars.
Elements Ne to Ca show a production factor which systematically reduces with the atomic number Z with also
large deviations from the O level: for example the block of elements Ne to Al are overproduced by a factor
of 3-5 with respect to O. If this were correct, it would automatically mean that O would not be mainly 
produced by massive stars (obviously if one assumes the elements from Ne to Al to be produced by massive stars).
The elements from Sc to Ni are all under-overabundant with respect to O (but the Co): this result is in line 
with the idea that these elements probably come from the ejecta of a Type Ia Supernova. The only exception is Co
which has a $\rm [Co/O]\geq0$: if this were correct we would face the unpalatable situation that the bulk of the 
Iron peak nuclei would come from one kind of star (type Ia Supernova) while just one single element of this 
group, Co, would come from massive stars. 

In the $\rm C_{0.2}$ case the only element which is clearly a problem is Na, which is embarrassingly
overproduced with respect to O ($\rm [Na/O]\simeq0.3$). All the other elements are more or less lined up 
with the currently mostly accepted scenario: all the elements from O to Ca share a very similar production factor,
which means that they all come from massive stars. C and N are underproduced with respect to O so 
that other sources (AGB's and the like) may contribute to their synthesis.
The same occurs for the elements beyond Ca which are all underproduced with respect to Oxygen by a factor 3-4,
leaving wide room for a Type Ia contribution to the galactic enrichment.

We hence conclude that, if the $\rm 25~M_\odot$ may be considered the leading polluter of the interstellar medium
and if the solar chemical composition is the "reference" distribution, a low C abundance, of the order of 0.2 dex
by mass fraction, should be left by the He burning.
A result very similar to the present one was already obtained by Weaver and Woosley (1993), who computed a large 
set of evolutionary models over different values of the $\rm ^{12}C(\alpha,\gamma)^{16}O$: by comparing their
results to the solar distribution they concluded that the "correct" $\rm ^{12}C(\alpha,\gamma)^{16}O$ rate should be
of the order of 1.7 times the rate quoted by CF88. Note that the C abundance they obtain at the end of the He burning
in the $\rm 25~M_\odot$ by adopting their "best" rate is 0.18 dex, i.e. remarkably similar to our standard case.
However, we have shown that the final C abundance left by the He burning does {\it not} depend on the size of
the convective core only if its border remains constant in mass when the central He drops below $\rm \simeq 0.1$ dex;
if this were not the case, the final C abundance would strongly depend on the behavior of the convective core.
Since we do not feel confident to state that we can robustly model the behavior of the convective core, we prefer
to interpret both our and their results in terms of the C abundance left by the He burning rather than in terms of
an effective $\rm ^{12}C(\alpha,\gamma)^{16}O$ rate.

More light will be shed shortly on this topic by an European task force which is beginning a new measurement
of such a tough cross section. Stay tuned... 

\acknowledgments
G.I. gratefully acknowledges the support of the Director of the Astronomical Observatory of Capodimonte, Prof. Massimo Capaccioli.
A.C. thanks the Astronomical Observatory of Rome and its Director, Prof. Roberto Buonanno,
for his generous hospitality at Monteporzio Catone. This paper has been partially supported by the 
MURST Italian grant (COFIN98).

\end{document}